\documentclass[sigconf]{acmart}

\usepackage{subcaption}
\usepackage{xspace}
\usepackage{booktabs}
\usepackage{textcomp}
\usepackage{xcolor}

\AtBeginDocument{%
  }

\copyrightyear{2025}
\acmYear{2025}
\setcopyright{cc}
\setcctype{by}
\acmConference[UIST '25]{The 38th Annual ACM Symposium on User Interface Software and Technology}{September 28-October 1, 2025}{Busan, Republic of Korea}
\acmBooktitle{The 38th Annual ACM Symposium on User Interface Software and Technology (UIST '25), September 28-October 1, 2025, Busan, Republic of Korea}\acmDOI{10.1145/3746059.3747748}
\acmISBN{979-8-4007-2037-6/2025/09}

\newcommand{\newh}[1]{{\textcolor{blue}{}}}
\newcommand{\delh}[1]{{\textcolor{red}{}}}

\begin{document}


\title{\textsc{Sensible Agent}: A Framework for Unobtrusive Interaction with Proactive AR Agents}




\settopmatter{authorsperrow=4, printfolios=true}
\author{Geonsun Lee}
\authornote{This project was undertaken during Geonsun's and Nels's internship at Google.}
\orcid{0000-0001-9401-8559}
\affiliation{%
  \institution{University of Maryland}
  \city{College Park}
  \state{MD}
  \country{USA}
}

\author{Min Xia}
\orcid{}
\affiliation{%
  \institution{Google XR}
  \city{Mountain View}
  \state{CA}
  \country{USA}
}

\author{Nels Numan}
\orcid{0000-0003-2931-7653}
\authornotemark[1]
\affiliation{%
  \institution{University College London}
  \city{London}
  \country{UK}
}

\author{Xun Qian}
\orcid{0000-0003-1976-7992}
\affiliation{%
  \institution{Google XR Labs}
  \city{Mountain View}
  \state{CA}
  \country{USA}
}

\author{David Li}
\orcid{0000-0002-3187-6190}
\affiliation{%
  \institution{Google XR Labs}
  \city{Mountain View}
  \state{CA}
  \country{USA}
}

\author{Yanhe Chen}
\orcid{0009-0007-6060-0410}
\affiliation{%
  \institution{Google XR Labs}
  \city{Mountain View}
  \state{CA}
  \country{USA}
}

\author{Achin Kulshrestha}
\orcid{0009-0004-8632-9575}
\affiliation{%
  \institution{Google XR}
  \city{Toronto}
  \state{ON}
  \country{Canada}
}

\author{Ishan Chatterjee}
\orcid{0000-0002-2123-6392}
\affiliation{%
  \institution{Google XR}
  \city{Seattle}
  \state{WA}
  \country{USA}
}

\author{Yinda Zhang}
\orcid{0000-0001-5386-8872}
\affiliation{%
  \institution{Google XR}
  \city{Mountain View}
  \state{CA}
  \country{USA}
}

\author{Dinesh Manocha}
\orcid{0000-0001-7047-9801}
\affiliation{%
  \institution{University of Maryland}
  \city{College Park}
  \state{MD}
  \country{USA}
}

\author{David Kim}
\orcid{0000-0002-0508-3509}
\affiliation{%
  \institution{Google XR Labs}
  \city{Zurich}
  \country{Switzerland}
}

\author{Ruofei Du}
\orcid{0000-0003-2471-9776}
\authornote{Corresponding author: Ruofei Du, me [at] duruofei [dot] com; Also contact: Geonsun Lee, gsunlee [at] umd [dot] edu}
\affiliation{%
  \institution{Google XR Labs}
  \city{San Francisco}
  \state{CA}
  \country{USA}
}

\renewcommand{\shortauthors}{Lee et al.}
\begin{abstract}
Proactive AR agents promise context-aware assistance, but their interactions often rely on explicit voice prompts or responses, which can be disruptive or socially awkward. We introduce \textsc{Sensible Agent}, a framework designed for unobtrusive interaction with these proactive agents. \textsc{Sensible Agent} dynamically adapts both “what” assistance to offer and, crucially, “how” to deliver it, based on real-time multimodal context sensing. Informed by an expert workshop (n=12) and a data annotation study (n=40), the framework leverages egocentric cameras, multimodal sensing, and Large Multimodal Models (LMMs) to infer context and suggest appropriate actions delivered via minimally intrusive interaction modes. We demonstrate our prototype on an XR headset through a user study (n=10) in both AR and VR scenarios. Results indicate that \textsc{Sensible Agent} significantly reduces perceived interaction effort compared to voice-prompted baseline, while maintaining high usability and achieving higher preference.

\end{abstract}



%

\begin{teaserfigure}
  \includegraphics[width=\textwidth]{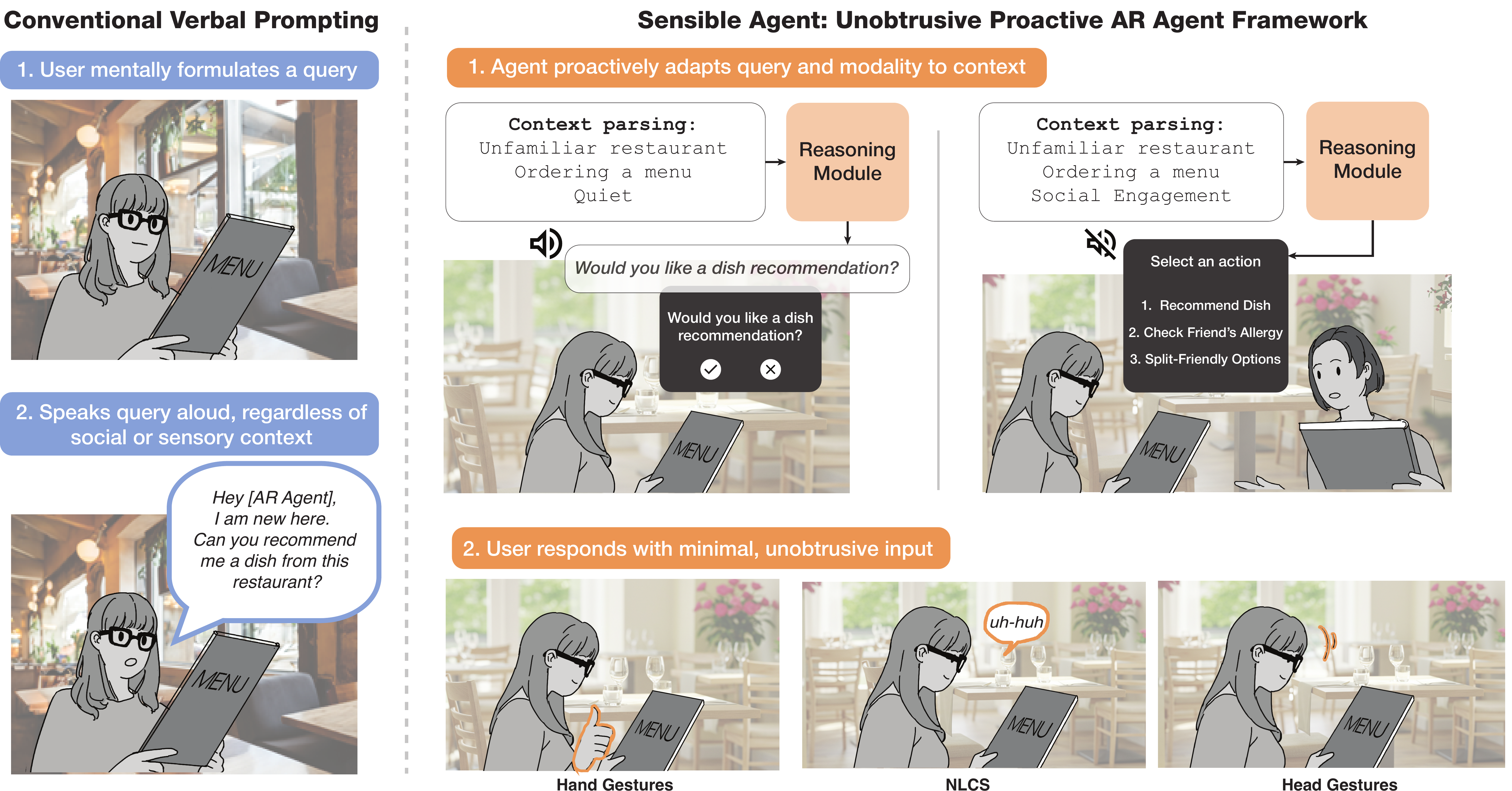}
   \vspace{-2.5em}
  \caption{We introduce \textsc{Sensible Agent}, a framework for unobtrusive interaction with a proactive AR agent. While the conventional approach requires users to use voice prompts to instruct agents, \textsc{Sensible Agent} proactively prompts the user based on context, toggles context-adaptive unobtrusive interactions, and suggests different types of queries based on the context.}
  \label{fig:teaser}
\end{teaserfigure}
\definecolor{whatblue}{HTML}{eaf6fc} 
\definecolor{howorange}{HTML}{ffe9d7}
\definecolor{lightgrey}{HTML}{E6E7E9}
\newcommand{\etal}{{\em et al.}} 
\newcommand{\eg}{{\em e.g.}} 
\newcommand{\ie}{{\em i.e.}}

\newcommand{\whatbox}[1]{\colorbox{whatblue}{\textcolor{black}{\small\texttt{\textbf{#1}}}}}
\newcommand{\howbox}[1]{\colorbox{howorange}{\textcolor{black}{\small\texttt{\textbf{#1}}}}}
\newcommand{\greybox}[1]{\colorbox{lightgrey}{\textcolor{black}{\small\texttt{\textbf{#1}}}}}

\maketitle

\section{Introduction}
As augmented reality (AR) and extended reality (XR) technologies become increasingly embedded in everyday life via smart glasses and head-mounted displays, researchers are re-imagining the roles of AI agents. Going beyond simply reacting to users' queries, these agents are envisioned as \textit{proactive assistants}, capable of anticipating and responding to user needs in situ~\cite{Liu2024Human,li2024omniactions}. Users often experience moments of sensory or cognitive disruption, particularly when navigating unfamiliar transit hubs or coordinating actions in crowded public spaces, challenging traditional input-driven interfaces. These moments call for agents that can assess user context and initiate timely support—even when the user’s hands, eyes, or voice are unavailable, thereby minimizing the need for explicit interaction.

Recent work has begun to explore proactive agents that surface knowledge or suggestions without direct queries. For instance, AiGet~\cite{cai2025aiget} utilizes smart glasses and large multimodal models (LMMs) to deliver incidental knowledge as users explore the world. However, AiGet is primarily designed for curiosity-driven engagement in low-stakes, slow-paced environments. In contrast, many real-world AR use cases involve time-sensitive, socially constrained, or cognitively demanding scenarios where the relevance and delivery of proactive assistance must be carefully tuned. In these settings, the question is not just \textit{what} to suggest, but \textit{whether}, \textit{when}, and \textit{how} it should be delivered. \autoref{fig:teaser} illustrates a motivating example of this contrast: whereas conventional agents rely on explicit voice commands in public, Sensible Agent unobtrusively adapts its prompts and input modalities based on real-time context.

Through a formative study involving 40 participants and 960 context-varying scenarios, we found that user preferences for both content and delivery modality vary widely depending on factors like temporal urgency, environmental sensory load, social presence, and task familiarity. For example, users in loud or crowded environments preferred subtle visual summaries over speech, while those in solo, focused settings accepted more direct audio prompts. These findings highlight the need for systems that treat proactivity as a context-sensitive coordination problem across intent, modality, and timing.

Prior systems address individual parts of this challenge. OmniActions~\cite{li2024omniactions} models user activity and context to recommend follow-up actions using LLM-based reasoning, but focuses intent prediction and digital workflow support rather than situated delivery. Human I/O~\cite{Liu2024Human} analyzes situational impairments, using multimodal sensing to dynamically adapt interaction modalities, but it does not address proactive intent generation or task-level reasoning. Existing AR interaction techniques often hardcode modality mappings or rely solely on environmental triggers alone~\cite{kumar2018use,vazquez2017serendipitous,weerasinghe2022vocabulary,caetano2023arlang}, limiting their responsiveness to nuanced shifts in user availability, attention, and social context.
Instead, we argue that, to be effective, proactive agents must jointly infer both \textit{what} to suggest and \textit{how} to present it, grounded in real-time user context and attentional load.
We argue that reasoning jointly over \textit{what} to do and \textit{how} to do it is not simply additive but essential for context-aware AR agents. A well-chosen modality cannot salvage an irrelevant suggestion, and a helpful prompt may go unnoticed if delivered via an ill-suited channel. This integration is particularly critical in socially sensitive or attention-limited settings.

In this paper, we present \textbf{\textsc{Sensible Agent}}, a context-aware proactive AR framework that adapts both the content (\textit{what}) and modality (\textit{how}) of its interventions. The framework comprises two modules: (1) an \textit{action suggestion module} using few-shot and chain-of-thought prompting with LMMs to recommend context-relevant agent actions, and (2) a \textit{modality selection module} determining the most suitable delivery channel (audio, visual, gestural, or passive) based on real-time multimodal input like gaze, ambient noise, hand availability. Both modules are guided by a taxonomy of action categories and context variants derived from our study, enabling structured conditioning of LLM outputs and policy decisions.

\textsc{Sensible Agent} is implemented as a WebXR-based prototype and evaluated on a diversity of realistic, daily scenarios presented via an Android XR headset. Our evaluation includes: (1) quantitative annotation of LLM outputs for action and modality appropriateness, (2) real-time latency benchmarking, and (3) scenario-based system demonstrations. We find that our dual-pipeline framework enables proactive support that aligns more closely with user expectations and contextual constraints than single-stream or modality-agnostic alternatives.

In summary, we contribute:
\vspace{-0.5em}
\begin{itemize}
    \item \textbf{\textsc{Sensible Agent}}, a context-aware, proactive AR agent framework that minimizes user interaction effort by jointly determining what to suggest and how to deliver it.
    \item \textbf{A user-derived design implications} of \textit{proactive actions} and \textit{context variants}, collected from a workshop study and a data collection study with 960 user responses across six everyday activities.
    \item \textbf{A functional prototype} using WebXR, multi-modal sensory input (e.g., hand gestures, head gestures, verbal command), and LLM prompting to demonstrate modality- and content-adaptive prompting behavior.
    \item \textbf{A user evaluation} (n=10) quantitatively and qualitatively comparing agent outputs and delivery strategies during a variety of realistic scenarios, presented via XR headset. 
\end{itemize}

Through this work, we take a step toward proactive AR systems that assist unobtrusively, adapt to real-world social and sensory contexts, and reduce the burden of communicating with intelligent agents.

\section{Related Work}
This section reviews prior work in context-aware AR, proactive agents, and multimodal interaction techniques. These areas collectively inform the design of our framework for unobtrusive, context-sensitive AR assistance.

\subsection{Context-Aware AR}
Context-aware AR---the ability to perceive and intelligently respond to environmental cues like spatial layout, objects, conditions, and user activity---is fundamental for creating effective and immersive experiences.
Recent advances in LLMs have significantly accelerated the capabilities of context-aware AR by enabling richer environmental understanding and more sophisticated spatial reasoning. For instance, Yang et al.~\cite{yang2024thinking} conducted a comparative study of Vision-Language Models (VLMs) to evaluate their spatial reasoning capabilities, providing valuable insights for AR applications. Furthermore, XaiR~\cite{srinidhi2024xair}, developed by Srinidhi et al., bridges the gap between large multimodal models and XR applications, while Xu et al.'s XAIR framework for multimodal 3D fusion and in-situ learning enables more sophisticated, spatially aware AI interactions~\cite{xu2024multimodal}.

Context-aware AR applications are emerging across various domains. Examples include AI cooking assistants in AR~\cite{jaber2024cooking,lee2024cookar}, semantic enhancement of object interaction~\cite{henderson2008opportunistic,Du2022Opportunistic,dogan2024augmented}, dynamic interface adaptation to context~\cite{li2024situationadapt}, and gaze-based and gesture-based disambiguation~\cite{lee2024gazepointar}. While these AR applications advance interaction with AI, they generally require explicit user input. In contrast, we study unobtrusive, proactive AR, where the system anticipates user needs and offers assistance that requires effortless user interaction.

Commercial XR systems like Apple Vision Pro\footnote{Apple Vision Pro: \url{https://www.apple.com/apple-vision-pro/}} and Meta Quest Pro\footnote{Meta Quest Pro: \url{https://www.meta.com/quest/quest-pro/}} offer rich multimodal input such as gaze, gestures, and voice, but their agents remain reactive, relying on user-initiated commands. These systems lack proactive agent behaviors that are contextually modulated based on user state, sensory availability, or social setting. In contrast, \textsc{Sensible Agent} integrates proactive content suggestion with adaptive modality selection, enabling agents to intervene in a timely and socially appropriate manner without requiring explicit user input.

\subsection{Agents and Proactivity}

Proactive agents aim to enhance user experience by anticipating needs and initiating interactions rather than solely reacting to explicit requests~\cite{liu2020towards,moon2019opendialkg,wu2019proactive,zhang2021dgpf,zhu2021proactive}. As outlined in a tutorial paper~\cite{liao2023proactive}, these proactive behaviors include learning to ask~\cite{bi2021asking,das2022automatic,li2024learning,ren2021learning,walker1995mixed,ye2022structured,zou2020towards,zou2019learning}, topic shifting~\cite{konigari2021topic,liao2020topic,tang2019target,xu2020conversational} and strategy planning with reinforcement learning, counterfactual dialogue act, and label generation~\cite{baek2022agreement,chen2020balancing,li2024learning,miller2021accuracy,tang2021high,tishby2000information}. For implementing proactive behaviors, multiple choice question answer~\cite{ren2023robots} allows us to define the problem as next-token prediction aligning well with LLM loss functions and training data.

    Beyond these core behaviors, various systems demonstrate proactive capabilities in specific contexts. Parse-Ego4D~\cite{abreu2024parse} offers personal action recommendation annotations. Satori~\cite{li2025satori} proactively guides users by modeling their mental states and environmental context in AR. YETI~\cite{bandyopadhyay2025yeti} learns scene understanding for potential intervention. Less or More~\cite{wang2025less} presents glanceable LLM explanations on smartwatches. COWPILOT supports web navigation~\cite{huq2025cowpilot} by suggesting next steps users can take. Similarly, OmniActions~\cite{li2024omniactions} predicts and suggests users action based on multimodal sensory inputs, such as images and audio. The system is triggered by certain actions such as scanning text or the event of taking a picture. 
    While OmniActions focuses on predicting potential user follow-up actions primarily for digital workflow support, our work centers on the challenge of \textit{situated delivery}. Sensible Agent adapts not only “what” assistance to offer, but crucially “how” to present it via \textit{unobtrusive} modalities selected based on real-time multimodal context sensing.

Other research explores proactive engagement for specific user needs. ComPeer is a text-based conversational agent that actively pings users based on previous conversation data and context to provide companionship and mental support~\cite{liu2024compeer}. Needs Companion defines a data model for service needs and uses a VA and LLM for needs elicitation and analysis through voice dialogue~\cite{nakata2024needs}. Zhang et al. leverages generative agents in a role-playing game to guide users to follow certain actions and in consequence elicit behavioral change such as environment-friendly behaviors~\cite{zhang2024can}. However, these works primarily focus on proactively guiding or intervening with users. In contrast, we center on minimizing interaction friction by adapting “what” assistance to offer and, crucially, “how” to interact based on real-time multimodal context sensing.


\subsection{Feedback Channels in Human-Agent Interaction}
Effective communication in human-agent interaction relies not only on the agent's output but also on the user's ability to provide appropriate, timely feedback. Prior research has explored a range of modalities through which users can signal feedback, ranging from explicit, intentional utterances to subtle paralinguistic and non-verbal cues.

\subsubsection{Explicit feedback}
Explicit input methods, such as spoken or typed commands, remain the primary means by which users provide feedback to voice-based agents. \citet{diederich2022design} identify communication modality—voice, text, or both—as a central design dimension of conversational agents, enabling users to issue requests, confirmations, or corrections in natural language. \citet{seymour2021exploring} highlighted the impact of speech as an interaction affordance and described how the shift to conversational interfaces has made interactions with assistants more social in nature.

\subsubsection{Whispering}
Whispering has emerged as a modality that enables private or low-disruption interactions with voice assistants. \citet{cho2019hey} examined how whispering affects user perceptions when querying sensitive health information. They found that whispering increased perceptions of social presence and comfort under low-sensitivity conditions. \citet{rekimoto2022dualvoice} introduced DualVoice, a whisper-classification mechanism that distinguishes between whisper and normal speech to support mode switching, e.g., whispering for commands and speaking normally for content input. In a follow-up work, the same author introduced WESPER~\cite{rekimoto2023wesper}, a system capable of converting whispered speech into audible speech in real-time, enabling silent, unobtrusive interactions in public spaces.

\subsubsection{Paralinguistic feedback}
Paralinguistic feedback includes non-lexical conversational sounds (NLCS), such as ``\textit{mm-hm}'', ``\textit{uh-huh}'', and ``\textit{oh}'', which convey a range of social cues. \citet{ward2006non} identified several categories of such sounds including acknowledgements, affirmation, disagreement, hesitation, and realization. These cues can express user feedback implicitly, signaling engagement, confusion, or alignment and can serve as an unobtrusive input method for users to provide feedback.




\subsubsection{Non-verbal feedback}
Emerging interaction contexts—such as walking, multitasking, or using devices in public—necessitate alternative input modalities beyond traditional hand or voice input. For instance, Cho et al.~\cite{cho2019hey} examined how users whisper to voice assistants when discussing sensitive health topics, finding that whispering fosters a greater sense of privacy and comfort. Here, we can see that the role of modality is not only in enabling interaction but also in shaping the social acceptability of agent use across contexts.

While walking, traditional interaction techniques such as hand gestures can be unreliable. Zhou et al.~\cite{zhou2016interaction} found that pinch gestures take significantly longer to execute when users are in motion, and hand input may be unavailable entirely when users are carrying items or wearing gloves. To address this, researchers have investigated hands-free input methods that are less sensitive to body posture or hand availability. These include intraoral input using tongue or lip movement \cite{jingu2023lipio}, and silent speech recognition through capacitive dental interfaces~\cite{kimura2022silentspeller}, though these approaches often require specialized hardware.

Other techniques leverage full-body motion during ambulation. Gaze input has been explored for hands-free cursor control, but may divert attention from environmental hazards, introducing safety concerns. In response, several works have explored interaction strategies tailored for mobile AR use. Lages and Bowman~\cite{lages2019walking} proposed interface adaptation techniques for AR transitions during walking. Müller et al.~\cite{muller2020walk} introduced WalkType, which maps lateral walking shifts to interface selection by rendering options as parallel paths on the ground. Kumar et al.~\cite{kumar2022passwalk} extended this technique by combining footpath gestures with gaze for secure AR headset authentication. More recently, GaitGestures~\cite{tsai2024gait} demonstrated that intentional changes in stride length and foot strike can be used as a low-effort, hands-free input method during locomotion.

Head-based input has also been explored as a socially acceptable and minimally disruptive modality. Tanenbaum et al.~\cite{tanenbaum2020make} use head rotation to control avatar facial expressions, while commercial systems like AirPods Pro incorporate simple head gestures for call handling. Prior research has also shown that nodding or pointing the nose can enable discrete UI selection\cite{kjeldsen2001head} and even serve as an authentication mechanism~\cite{li2016whose}, making head gestures a viable modality for subtle, context-aware interactions in AR.

\subsection{Interacting with AR in public}
Integrating AR devices into public settings presents unique challenges concerning user comfort and social acceptability. Addressing these factors is crucial for facilitating seamless and comfortable public interactions with AR technologies.

Recent studies have explored the social dynamics of AR usage in communal environments. For instance, Kaeder et al.~\cite{kaeder2024working} investigated how different virtual display layouts affect users’ perceived productivity, feelings of safety, and social acceptability when working with mixed reality in public spaces.

Similarly, Pavanatto et al.~\cite{pavanatto2024working} examined both user and bystander experiences of XR displays in real-world settings. The study revealed that while users generally accept XR technology in public, factors such as previous XR experience and personality traits can impact perceptions.

Lu and Bowman~\cite{lu2021evaluating} introduced the concept of Glanceable AR interfaces, designed to provide users with quick, unobtrusive access to information through peripheral displays. Their in-the-wild evaluations demonstrated that such interfaces are less distracting and more socially acceptable for everyday tasks in public settings.
Incorporating these insights, our framework emphasizes the development of AR interactions that are not only functional but also socially considerate. By focusing on user-centric design principles, we aim to facilitate AR experiences that users can comfortably and confidently engage with in public settings.


Informed by this prior work on feedback channels, we introduce Sensible Agent. Our framework focuses on unobtrusive proactivity by dynamically adapting both “how” assistance is delivered and “what” feedback modalities are supported, based on real-time multimodal context.

\section{Workshop Study}
To explore users' motivations and scenarios in which they would require a proactive AR agent, as well as their preferred interaction methods, we conducted a workshop study.

\subsection{Study Procedure}
We recruited 12 participants internally from Google with diverse backgrounds (engineers, designers, researchers, students). The workshop began by introducing proactive AR agents, contrasting them with the existing user-prompted AR agents (illustrated via a Project Astra~\cite{GoogleDeepMind_ProjectAstra_2024} video). Using a shared digital whiteboard, participants brainstormed over two structured ideation rounds, each addressing a specific research question:

\begin{itemize}
    \item \textbf{RQ1:} What types of proactive queries would users like the agent to initiate? Specifically, in what \textit{situation} should the agent act, what \textit{action} or \textit{query} should it perform, and \textit{why} is proactive behavior necessary?
    \item \textbf{RQ2:} How should users interact with the agent in public settings? This included considerations of both the \textit{output modality} (how the agent should present its proactive queries) and the \textit{input modality} (how users would respond).
\end{itemize}

In each round, participants had 10 minutes to reflect and post ideas, followed by group discussion to present and aggregate thoughts. Two moderators ensured equal participation and captured key points. The workshop lasted approximately one hour.

\subsection{Findings}
We conducted a thematic analysis of responses, drawing on design space frameworks from prior art~\cite{maclean2020questions,xu2023xair}. We focused on contexts where users desired proactivity, the actions they expected, and their reasoning. These findings shaped our framework's \textit{``what''} and \textit{``how''} modules.
These findings shaped the design of our framework, particularly how the agent determines \textit{``what''} to do and \textit{``how''} to present it.
Participants are denoted as W1–W12.

\subsubsection{\textbf{When} do we need proactive AR agents?\\} 
Participants described scenarios in which they expected agents to anticipate needs and act without explicit input. Across 12 participants and 45 scenarios, six recurring contextual factors emerged:

\textbf{Repetitive, Predictable Activities (n=9).} Participants described routine scenarios in which their next actions were both predictable and repetitive. They expressed frustration with having to repeatedly issue explicit commands for tasks they perform frequently. For instance, W1 noted, \textit{“I have daily routines such as taking a bus and playing Spotify. In this case, the agent should learn this and suggest it first when I get on the bus”}.

\textbf{Public or Socially Awkward Situations (n=6).} Participants noted that verbal interaction was socially uncomfortable or inappropriate in public or quiet environments (\textit{\eg}, libraries, cafes).

\textbf{Uncertainty or Lack of Awareness (n=4).} Participants described situations in which they were unsure what assistance to request or unaware of the agent's available capabilities. In these cases, proactive suggestions were seen as beneficial. W9 mentioned: \textit{“Sometimes, I don't even know what's possible to ask; proactive suggestions would help me discover useful actions.”}

\textbf{Unfamiliar Environments or Activities (n=5).} Participants desired proactive guidance in new settings or during unfamiliar activities, such as visiting a new city or starting a hobby. Some emphasized the need for varied suggestions in these situations (W6, W7, W10).

\textbf{Time-sensitive Scenarios (n=4).}
Participants identified high-pressure contexts (\textit{\eg}, rushing to catch transport) as prime opportunities for proactive assistance. W5 said: \textit{“When I'm rushing, I don't have the mental capacity to have a full blown conversation with an agent.”}

These findings illustrate the necessity of considering contextual variables like familiarity, social environment, urgency, and uncertainty to effectively shape \textbf{\textit{what}} proactive actions the agent should suggest.

\subsubsection{\textbf{What} do we want proactive AR agents to do?\\}
We found significant overlap between the contexts of \textbf{\textit{when}} and participants' suggestions for \textbf{\textit{what}} the agent should do. However, participants also explicitly articulated desired proactive actions:

\textbf{Information Delivery (n=11).} Delivering relevant context-aware information without explicit user query (\eg, translating menus, recognizing landmarks, or offering pronunciation feedback in language learning).

\textbf{Reminders and Notifications (n=9).} Nudging users about forgotten intentions or events based on routine or temporal triggers (\eg, picking up medication, sending messages when late, stocking household items).

\textbf{Suggestion and Option Surfacing (n=6).} Offering creative or exploratory ideas when users have no concrete goals (\eg, suggesting interior decor, restaurant options, or AR visualizations for artwork).

\textbf{Error Detection and Guidance in Tasks (n=5).} Providing real-time guidance during procedural or skill-based tasks when users pause, struggle, or deviate (\eg, correcting instrument fingering, helping with furniture assembly, pointing out cooking errors).

\textbf{Environment or Object Control (n=4).} Automatically interacting with physical or digital systems based on routine or context (\eg, turning off lights, logging food, muting phone calls while driving).

While specific actions varied, participants consistently preferred contextually timed, non-intrusive suggestions that reduced the burden of remembering, navigating UIs, or formulating questions.

\subsubsection{\textbf{How} do we want to interact with proactive AR agents?}
Participants emphasized the need for unobtrusive, contextually appropriate interaction methods:

\textbf{Hand Gestures (n=10).}
Participants frequently suggested hand gestures as subtle means of interaction but explicitly noted limitations when engaged in hand-intensive tasks.

\textbf{Head Gestures (n=5).} Simple head movements such as nodding, shaking, or slight tilting were popular due to their subtlety and intuitive nature. \textit{``If the agent asks a simple yes-no question, I could just slightly nod or shake my head without anyone noticing,”} explained a participant (W7).

\textbf{Gaze-based Interactions (n=5).}
Gaze inputs like blinking or gaze-dwelling were identified as useful, especially when hands were unavailable or the user wished to interact privately.

\textbf{Subtle Auditory Inputs (n=4).}
Participants suggested NLCS and whipsering as an alternative for a less intrusive way than overt speech.

\textbf{Integrated Activities (n=3).}
Some participants proposed embedding response to proactive suggestions into ongoing activities (\eg, continuing a task as implicit confirmation).

Nine participants explicitly or implicitly referred to Situationally Induced Impairments and Disabilities (SIIDs)~\cite{wobbrock2019situationally,Liu2024Human}, which influenced their preferences for both how the agent should present itself and how they would interact with it.

\subsection{Design Implications}

Our findings suggest that users' expectations for proactive AR agents are shaped not only by the activity at hand but also by fine-grained situational factors that influence the relevance of proactive actions (\textit{what}) and the appropriateness of interaction modalities (\textit{how}). We outline five key design implications that directly informed our framework:

\textbf{(1) The same activity may require different proactive behaviors based on contextual variants.}  
Participants often described repeated tasks (\eg, navigating an airport, visiting a museum) where the proactive action varied depending on environmental familiarity, time pressure, or social setting. This suggests that static task-based modeling is insufficient. Proactive systems must account for how variations in context shift the user's expectations. Our framework addresses this through a context similarity module in the \textit{what} pipeline, allowing the system to adapt actions to nuanced differences across similar scenarios.

\textbf{(2) Social engagement and public settings significantly constrain interaction modalities.}  
Participants expressed reluctance to interact with agents via speech or overt gestures in socially sensitive environments (\eg, meetings, public transit). This highlights the need to reason about social acceptability—not just sensor availability—when choosing output and input modalities. In our framework, this is addressed by incorporating social engagement as an explicit factor in determining interaction modality.

\textbf{(3) Temporarily impaired input/output channels are common in everyday settings.}  
Rather than permanent disabilities, participants frequently described moments where they were visually, audibly, or physically unavailable due to the activity (\eg, eating, driving, holding an object). These temporary impairments---also described in prior work such as SSID~\cite{Liu2024Human}---should be treated as first-class input to the system. Our framework integrates this insight into the \textit{how} module, enabling dynamic modality selection based on real-time multimodal sensing.

\textbf{(4) Users welcome suggestion diversity when uncertain, but prefer precision when familiar.}  
When participants were unsure of their goals or facing novel tasks (\eg, decorating a room, visiting a museum abroad), they welcomed diverse suggestions. In contrast, familiar routines called for focused, streamlined actions. In response to this, our system's suggestion generation module varies the breadth and structure of proactive queries (\eg, binary vs. multi-choice) depending on user familiarity with the activity.

\textbf{(5) Embedded, multimodal confirmations lower friction in high-effort scenarios.}  
Participants often preferred confirming or rejecting proactive suggestions through natural, low-effort behaviors (\eg, nodding, gaze dwelling, continuing the current task). This indicates that confirmation should be implicitly embedded in the interaction rather than handled through explicit follow-ups. Our interaction module prioritizes combining multimodal signals (head, hand, voice, gaze) to enable confirmation mechanisms with minimal cognitive or physical effort.

\section{\textsc{Sensible Agent}: A Framework for Context-aware Unobtrusive Proactive Agents}


We present \textbf{\textsc{Sensible Agent}}, a framework for building proactive AR agents that prioritize unobtrusive interaction and minimal user effort. Unlike traditional systems that rely on user-initiated queries, our framework is designed to anticipate user needs and respond proactively, while adapting both the content and delivery of suggestions to situational context.



\begin{figure*}[h]
  \centering
  \includegraphics[width=0.9\linewidth]{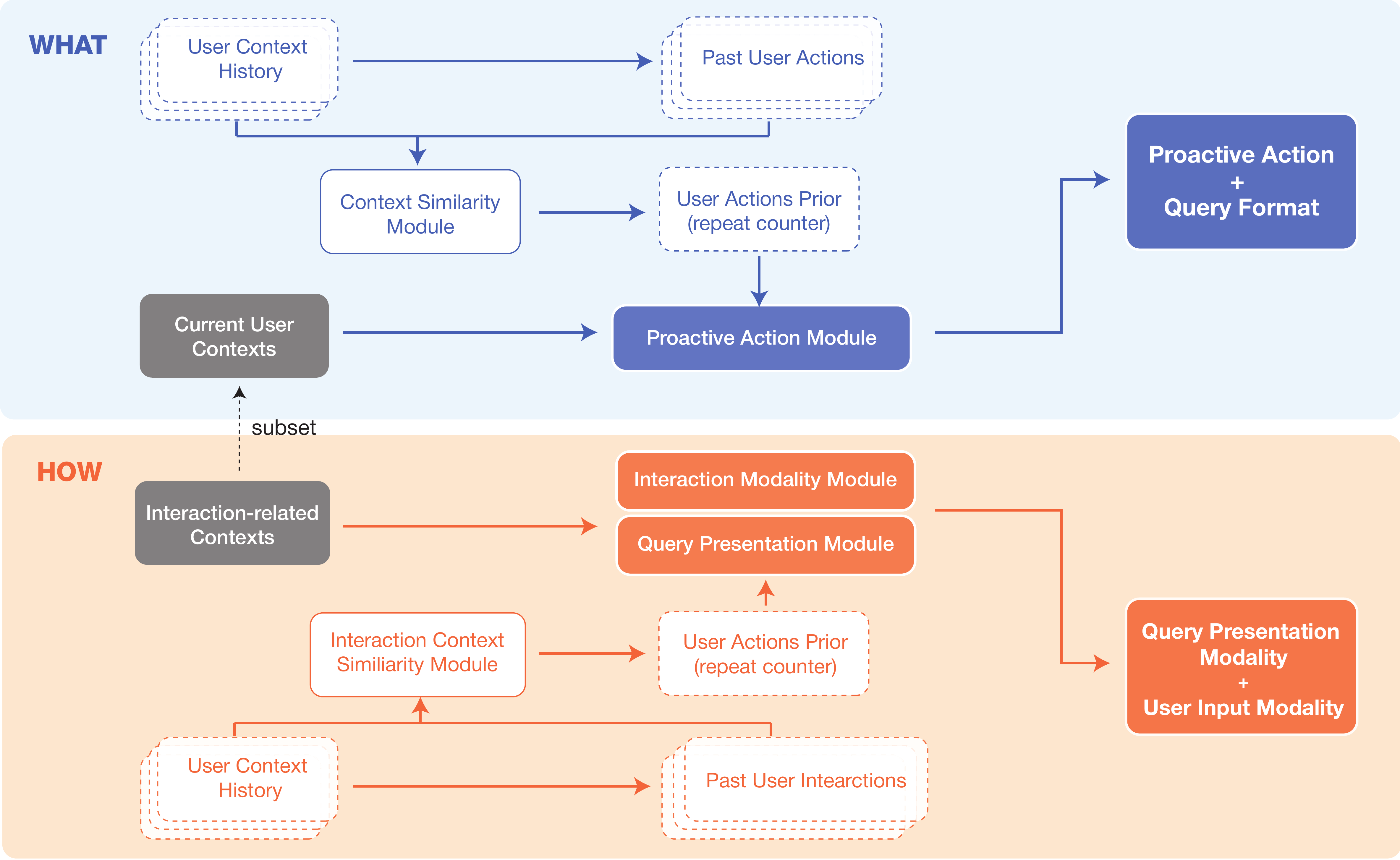}
  \caption{Detailed dataflow of the \textsc{Sensible Agent} framework. An \whatbox{ACTION RECOMMENDATION MODULE} (\whatbox{WHAT}) takes user context and determines the suggested action in one of three primary formats, and an \howbox{INTERACTION ADAPTION MODULE} (\howbox{HOW}) selects presentation modality and input modalities.}
  \label{fig:framework}
\end{figure*}

The framework consists of two interdependent reasoning modules, as illustrated in \autoref{fig:framework}: the \whatbox{ACTION RECOMMENDATION MODULE} (\textit{What}) and the \howbox{INTERACTION ADAPTION MODULE} (\textit{How}). While our prototype implements the core components of both modules, certain capabilities such as context similarity based on long-term user history are part of the envisioned design and discussed as future directions.

\subsection{Action Recommendation Module (What)}

This module determines what actions the agent should proactively suggest in a given context. It is designed to anticipate user intent and reduce decision-making burden by tailoring suggestions to situational cues and prior behavior.

\textbf{User Context Adaption.} The 
\small{\textbf{\texttt{CONTEXT SIMILARITY MODULE}}}\normalsize uses the \textit{current user context}, extracted in real time and encompassing dimensions such as activity and location familiarity, perceived cognitive load, social engagement, and temporal urgency. In the 
framework, it also draws on a \textit{user context history} and \textit{prior user actions}, enabling the system to learn from repeated patterns or behavioral regularities.



\textbf{LMM Reasoning.} Based on the current context—and user history of similar context—the \whatbox{PROACTIVE ACTION MODULE} not only determines \textit{what} action(s) to suggest but also selects the most appropriate \textit{presentation format}. We define three three primary formats, ordered from most to least effort for the user:

\begin{itemize}
    \item \textbf{Multi-choice Selection:} Presented when several possible actions may be contextually appropriate, allowing the user to choose from.
    \item \textbf{Binary Confirmation:} Employed when a single action is predicted with high confidence, but explicit user confirmation via a `yes'/`no' respond is required.
    \item \textbf{Icon-based Cue:} Deployed for highly probable, low-stakes actions where user intent is inferred with very high confidence. The agent proactively visualizes a relevant graphical icon (\eg, a translation/menu icon) in peripheral region, affording user interaction with minimal interruption. 
\end{itemize}


\subsection{Interaction Adaption Module (How)}

This module determines how the proactive suggestion should be delivered and how the user should interact with it, based on real-time input/output availability and context-driven appropriateness.

\textbf{User Context Adaption with I/O Channels Availability.} The \howbox{HOW} module shares the same core \textit{user context} extracted for the \whatbox{WHAT} module but further considers \textit{input-related context}, such as whether the user’s hands are occupied, their environment is noisy, or they are engaged in conversation. These additional cues reflect situationally induced constraints that affect interaction feasibility.

\textbf{User Input Similarity.} While not yet implemented, \small{\textbf{\texttt{INTERACTION}}} \small{\textbf{\texttt{CONTEXT SIMILARITY MODULE}}} \normalsize component would compare current input-related context to prior interaction patterns, helping refine modality decisions based on similarity to previously successful input conditions. This design is informed by the concept of SIIDs~\cite{Liu2024Human}, which describe temporary constraints on user input/output channels.

\textbf{Presentation Strategy.} The \howbox{QUERY PRESENTATION MODULE} selects how the agent’s proactive suggestion is presented to the user, choosing from:

\begin{figure*}[h]
  \centering
  \includegraphics[width=\linewidth]{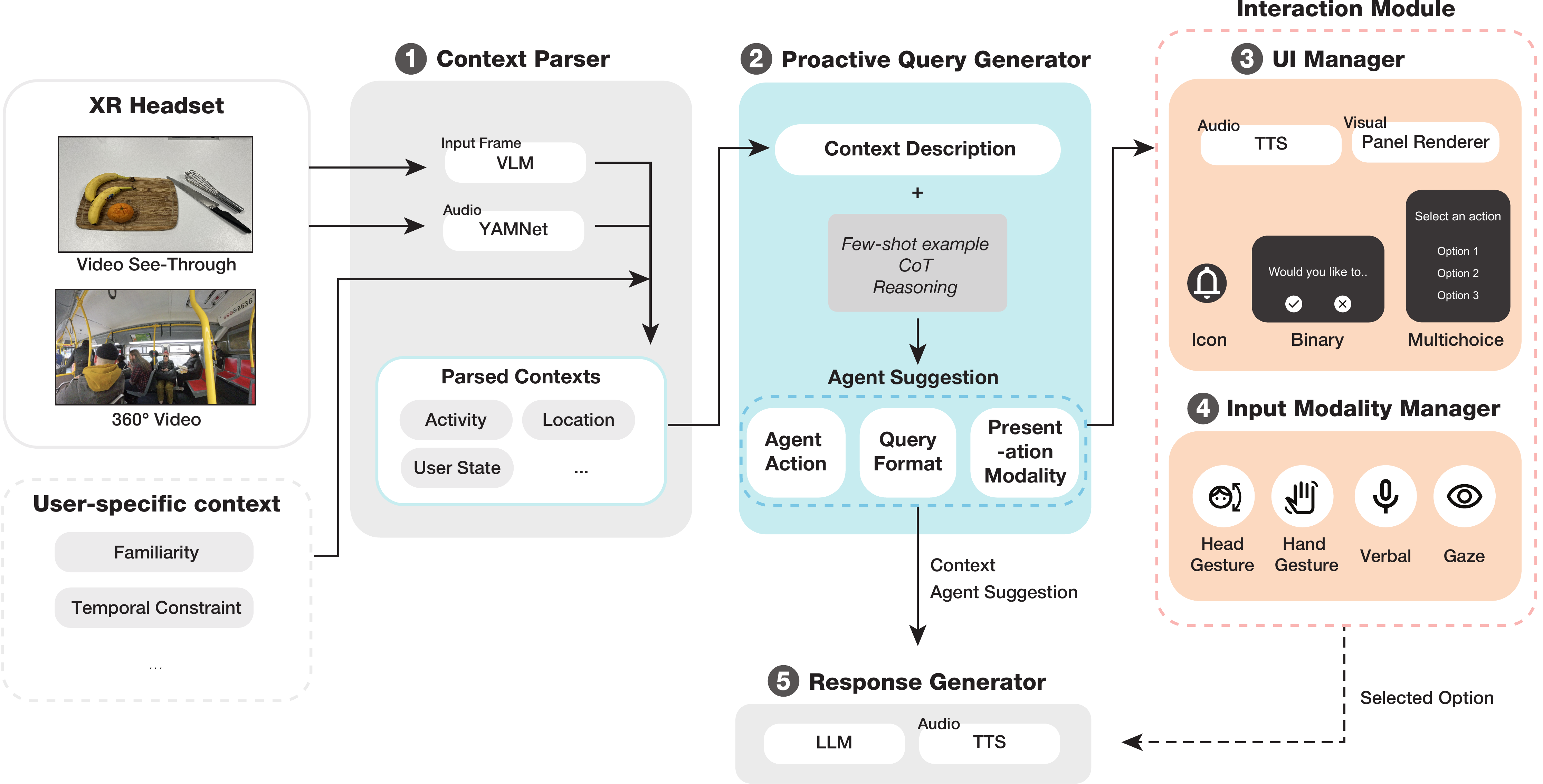}
  \caption{System architecture of our proactive AR agent prototype. The full system is implemented in WebXR with support for real-time interaction in 360$^{\circ}$ videos or video see-through AR environments. The system processes visual and audio input (1) and parses contextual attributes such as familiarity, urgency, and environmental noise using a VLM and YAMNet.  (2) Based on the parsed context, the proactive query generator formulates a suitable suggestion, including its agent action, presentation modality, and query type. These are passed to the interaction module, (3) where the UI manager renders the query and the (4) input modality manager enables one or more input modalities (\eg, gaze, hand, head, voice) based on feasibility and appropriateness. The interaction module then forwards the selected option by the user to the (5) response generator.}
  \label{fig:system_architecture}
\end{figure*}

\begin{itemize}
    \item \textbf{Visual-only:} On-screen UI elements, icons, or overlays.
    \item \textbf{Auditory-only:} Spoken messages or system voice prompts.
    \item \textbf{Audio-visual:} Redundant or complementary presentation across both channels.
\end{itemize}
Presentation strategy is determined based on environmental and social context. For example, in a quiet or very noisy public setting, the agent may suppress audio and rely on visuals, while in visually demanding tasks (\eg, biking), auditory prompts are prioritized.

\textbf{Interaction Modality Adaption.} The \howbox{INTERACTION MODALITY} \howbox{MODULE} determines which input modalities are enabled for confirming or responding to proactive suggestions. It considers both the user's input-related context and the selected presentation strategy. For instance, if the output is visual-only and the user is not looking at the screen, gaze-based input is not viable. Modalities supported include:

\begin{itemize}
    \item \textbf{Gaze (dwell):} Used for visual interfaces, enabling binary or multi-choice input by tracking where the user looks. Buttons are triggered by holding still for one second.
    \item \textbf{Hand gestures:} Uses explicit gesture recognition (\eg, open palm, fist) rather than raycast-based pointing, enabling confirmation or selection even when direct targeting is not feasible.
    \item \textbf{Head gestures:} Supports nodding and shaking for binary prompts, and directional tilting (\eg, left, right, backward) for multi-choice selection.
    \item \textbf{Voice input:} Enables spoken responses using lightweight verbal commands. For binary interactions, users may respond with naturalistic non-conversational lexical sounds (NCLS) such as ``\textit{uh-huh}'' or ``\textit{mmm-mm}.'' For multi-choice prompts, the agent supports one-word commands such as ``\textit{one},'' ``\textit{two},'' or ``\textit{three}.''
\end{itemize}

These input modalities can be used independently or in combination, depending on user availability and task constraints. The design prioritizes interaction methods that are socially acceptable and impose minimal cognitive or physical load.

\subsection{Module Integration}

The two modules operate in parallel and share context inputs. The \whatbox{WHAT} module determines the structure and content of the suggestion, which then informs the \howbox{HOW} module’s choice of presentation and interaction modality. For instance, a multi-choice suggestion during a high-cognitive-load task may trigger a visual interface with hand gesture input, whereas an icon-based prompt during a routine task may use gaze-only interaction.

Together, these modules support a context-aware proactive agent that adapts not only to what the user needs, but how they can most easily engage. In the following section, we describe how we implemented the core components of this framework in a functional prototype and outline the system capabilities currently supported.

\section{Prototype Implementation}\label{sec:implementation}
We implemented a WebXR-based working prototype of our context-aware proactive AR agent system, focusing on the core modules identified in our framework: the proactive action module and the adaptive interaction modality module. Our system leverages LMMs to infer real-time context and generate proactive suggestions and interaction strategies accordingly.

We describe the overall architecture and interaction flow of the system, followed by details of our data collection study, which informed the agent's suggestion generation logic.



\subsection{System Architecture}

\autoref{fig:system_architecture} illustrates the overall architecture of our prototype, including context parsing, query assembly, inference, and agent response integration.
The system operates on egocentric video input, which can come from a real-time video see-through (VST) stream through Android's Camera2 API\footnote{Camera2 API: \url{https://developer.android.com/media/camera/camera2}} or a 360\textdegree{} pre-recorded video. The latter is primarily used to simulate environmental contexts in controlled study conditions where live capture is not feasible.

When a trigger event is detected--such as a pause in activity or gaze fixation--a single image frame is extracted and sent to the system's reasoning pipeline. This pipeline consists of three layers: a context parsing layer, a proactive query generation layer, and a response generation layer. Each layer operates using LMMs (GPT-4o), conditioned through in-context learning with structured input examples derived from our data collection study.

\paragraph{Context Parsing.} The first layer uses the visual input along with manually injected user-specific variables, such as task familiarity or temporal urgency, to identify key contextual attributes. These include physical environment (\eg, type of space), user state (\eg, hands occupied, conversational setting), and social or environmental constraints (\eg, noise level, presence of others). This layer simulates the role of the context similarity module in our framework, although it does not perform explicit similarity computation across historical data. We use Chain-of-Thought (CoT) prompting here to extract structured outputs through intermediate reasoning steps, enabling more accurate context decomposition.

\paragraph{Proactive Query Generation.} The output from the first layer is passed to the second layer, which generates the proactive query the agent should present. This includes three elements: the action content (\eg, suggesting information or assistance), the query format (multi-choice, binary, or icon-based), and the suggested presentation modality (visual, auditory, or both). Although our framework initially distinguishes between the \whatbox{WHAT} and \howbox{HOW} modules, we found that the decision around presentation modality is closely entangled with the action content and context. As such, this component is computed together with the query in the second layer, while the input modality constraints are handled downstream.

\begin{figure*}[h]
  \centering
  \includegraphics[width=\linewidth]{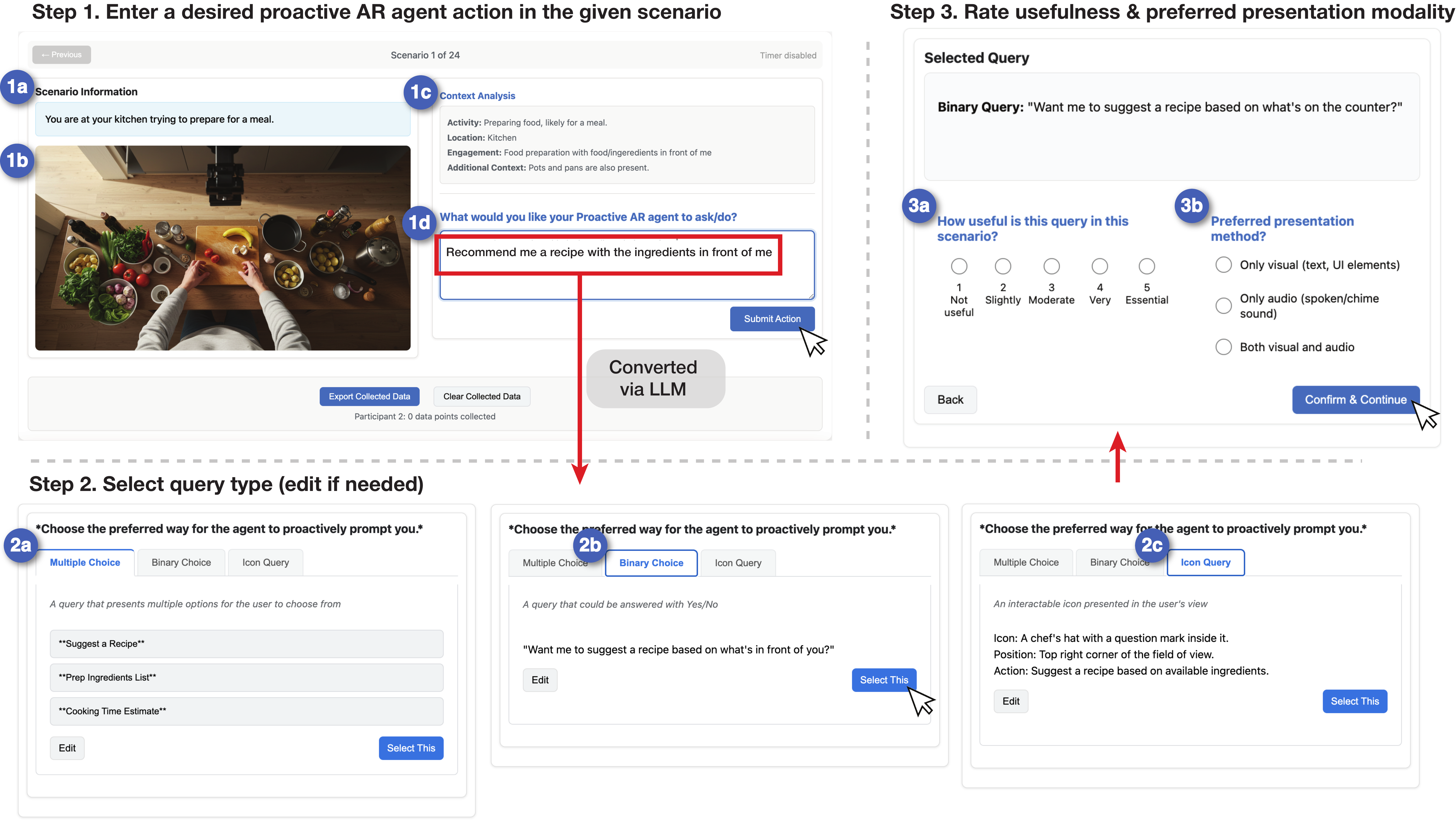}
  \vspace{-2em}
  \caption{
    Web interface for the data annotation study. Each participant annotated 24 scenarios through a 3-step workflow. In Step 1, participants viewed: \textbf{(1a)} a short text describing the scenario, \textbf{(1b)} a synthetic egocentric image for visual consistency, \textbf{(1c)} contextual details (\eg, location, engagement), and \textbf{(1d)} a text input field to describe the desired proactive AR agent action. In Step 2, the input was converted into \textbf{(2a)} multi-choice, \textbf{(2b)} binary, and \textbf{(2c)} icon-style queries using LLMs. Participants could edit or choose their preferred query type. In Step 3, they \textbf{(3a)} rated the usefulness of the action and \textbf{(3b)} selected the preferred presentation modality (audio, visual, or both). Final responses were exported as a CSV after completing all 24 scenarios.    }

  \label{fig:data_web}
\end{figure*}

\paragraph{Interaction Module.} The UI manager then constructs a panel interface or an audio playback using OpenAI's text-to-speech (TTS) API~\cite{openai_tts_2024} based on the generated suggestion and modality. At the same time, the parsed context is sent to the input modality manager, which enables a subset of input modalities. These include head gestures, hand gestures, gaze (via dwell), and verbal inputs. Each modality is gated based on two criteria: the inferred contextual appropriateness (\textit{\eg}, SIIDs~\cite{Liu2024Human}) and the feasibility given the chosen output modality . For example, if the context suggests that the user is in a noisy public setting, voice input is disabled and visual interactions such as gaze or head gestures are prioritized. Similarly, if a query is presented solely in audio form, gaze interaction is considered infeasible and suppressed.


\paragraph{Response Generation Layer.}
Once the user confirms or selects an option from the proactive prompt, the system passes both the structured context and the selected action to an LLM to generate a natural language response. This response serves as the agent’s follow-up behavior, grounded in the user's selection (e.g., providing details about a painting the user is viewing in a museum). The generated utterance is then synthesized via TTS and played back through the headset's audio channel.

\begin{figure*}[h]
  \centering
  \includegraphics[width=\linewidth]{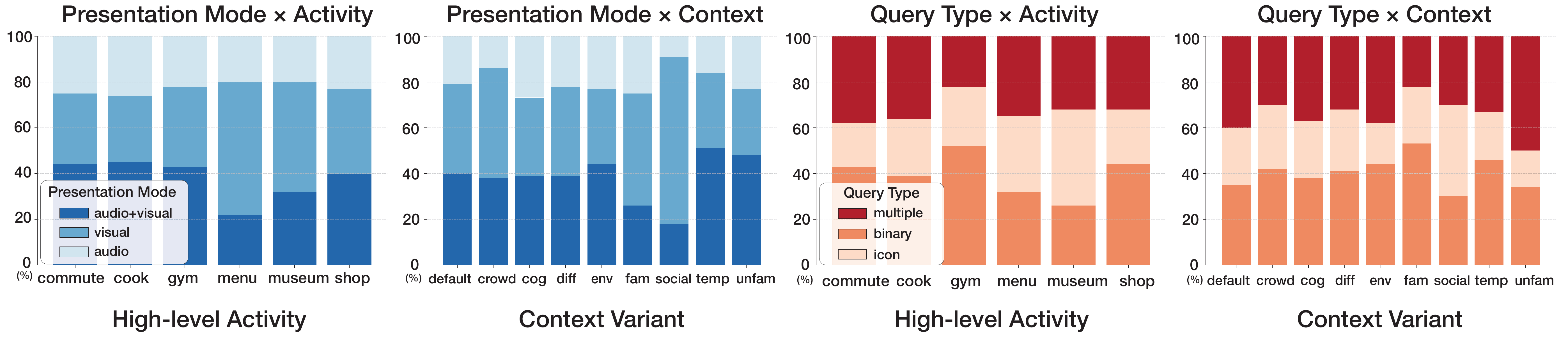}
  \vspace{-2.5em}
  \caption{
    Distribution of data entries in selected presentation modes (left) and query types (right) across different high-level activities and context variants. Data showed varying preferences for modality (audio, visual, audiovisual) and query format (binary, multiple-choice, icon-based) depending on situational demands and activity type.
    }

  \label{fig:dcs_data}
\end{figure*}

\subsection{Proactive Actions Data Collection Study}
{\label{sec:dcs}}

To support context-aware query generation in our proactive agent system, we conducted a data collection study designed to elicit how users' expectations of agent behavior vary based on situational context. This study informed the core reasoning mechanism of our system by providing grounded examples of context-action-modality mappings, which were later used to condition an LLM through in-context learning.

\paragraph{Web-Based Annotation Interface.}  We developed a custom web-based annotation interface (see Figure~\ref{fig:data_web}) that presents participants with egocentric-view scenarios. Each scenario consisted of a synthetic image paired with a one- to two-sentence description simulating an immersive AR context. To ensure interpretability and consistency, we also provided optional structured descriptors--such as location, detected high-level activity, user engagement, and environmental or social context--displayed in a collapsible panel. These served as objective reference points for participants, complementing the visual and narrative descriptions.

Participants were asked to imagine themselves in each situation and enter, in free-text form, what they would want a proactive AR agent to do or ask on their behalf. Upon submission, an LLM (GPT-4o)  was used to reformat the user-described action into three proactive query forms: multi-choice, binary, and icon-based. Participants could revise the generated phrasing and select the query form they felt was most appropriate for the given context. They were also asked to rate how useful they believed the proactive suggestion would be in the scenario on a five-point Likert scale, and to specify their preferred modality of presentation (audio, visual, or both). Participants could navigate between scenarios at any point and revise their annotations.

\paragraph{Presented Scenarios.} The study included 48 scenarios, each representing a variant of one of six high-level activities commonly encountered in daily life: reading a menu at a restaurant, working out at a gym, grocery shopping, browsing in a museum, commuting by public transportation, and cooking in a kitchen. Each activity contained 5 to 8 contextually distinct variants, which were designed to reflect factors such as location and activity familiarity, situational impairments (\eg, hands occupied), social constraints, and temporal urgency. Each participant annotated 24 scenario of three selected high-level activities, with the task taking approximately 30 to 40 minutes to complete.

\paragraph{Participants.} We recruited 40 participants through internal mailing lists and social platforms. Participants varied in age (21 - 44) and background and included individuals with prior experience using AR headsets or glasses (34 out of 40). On average, participants reported a $\mu = 3.97$ mean experience with AR and $\mu = 3.57$ with voice assistants on a 5-point Likert scale. 

\subsubsection{Results and Analysis}

We analyzed 960 query entries from 40 participants (24 per person) across six high-level activities and their contextual variants. Our analysis focuses on (1) consistency in query preferences, (2) variation in query type and presentation modality by context, and (3) a taxonomy of desired proactive actions. (\autoref{fig:dcs_data})

\textbf{Dataset Overview}
We analyzed a total of 937 proactive action entries from 40 participants across 48 scenarios. Each participant encountered scenarios drawn from six high-level activity types—menu reading, cooking, visiting a museum, commuting, working out, and grocery shopping—embedded in varied contextual conditions such as familiarity, temporal urgency, or social engagement. For each scenario, participants (1) described their desired agent action, (2) selected a preferred query type (binary, multi-choice, or icon), (3) chose a presentation modality (audio, visual, or audio+visual), and (4) rated the usefulness of the action out of a 5-point Likert scale. To understand preferences for \textit{how} the agent should assist, this analysis focuses on the 937 of 960 entries (97.6\%) where users desired proactive help (usefulness rating $> 1$).

\textbf{Query Format is Shaped by Contextual Demands}
Query type selection was highly context-sensitive. Overall, multi-choice (42\%) and binary queries (36\%) were more frequently preferred than icon-based options (22\%) across high-level activities. However, preferences shifted meaningfully across context variants.

In unfamiliar scenarios, users often favored multi-choice formats to explore alternative paths or receive richer input from the system (50\%)). For example, in an unfamiliar restaurant context, a participant selected a multi-choice query, \textit{“1.Translate dish names, 2.show images, 3.suggest the most popular dish,”} and rated it highly useful (5/5). On the other hand, binary queries were more common under temporal pressure or when rapid assistance was needed (48\%)), such as, \textit{“Would you like me to recite your grocery list?”}

Icon-based formats, though less common overall, emerged in socially sensitive (40\%)) or familiar environments (25\%) where minimal interaction was desired. In one such instance, a participant requested, \textit{“Show vegan options for my friend,”} during a socially-engaged dining setting, paired with an icon query and visual-only presentation.

These findings suggest that the agent’s querying mechanism should be sensitive to both task complexity and situational constraints, offering lower-friction formats in fast-paced or public-facing contexts while enabling richer interactions when users have time and attention to spare.

\textbf{Presentation Modality Vary by Contextual Constraints\\}
Across all scenarios, the majority of participants preferred audio-visual presentation (38\%), likely due to the redundancy and clarity it offers. However, this preference was not universal.

In socially dense or quiet public settings, such as museums or restaurants, users gravitated toward visual-only queries. For example, in a crowded museum, one user asked the agent to \textit{“Show artwork info I am looking at,”} choosing visual-only presentation for discretion. On the other hand, audio-visual modalities were favored in unfamiliar or time-constrained scenarios, where rapid and clear communication was necessary—\eg, \textit{“Suggest fast options I can eat from the menu.”}

These results reinforce the need for modality adaptation in agent design, modulated by situational context and the user's social environment.

\begin{table*}[t]
\centering

\footnotesize
\begin{tabular}{p{4.2cm} p{4.2cm} p{4.2cm} p{1.6cm} p{1.6cm}}
\toprule
\textbf{Context Description} & \textbf{CoT Reasoning} & \textbf{Agent Action} & \textbf{Query Type} & \textbf{Modality} \\
\midrule
User is in a museum, crowded with people and slightly noisy, while engaged with an art piece. &
User may not hear audio clearly and is visually focused on the artwork. A visual, low-effort query is ideal. &
Offer more information about the artwork (e.g., title, artist, background). &
Icon & Visual \\
\midrule
User is in a familiar grocery store but is in a rush, quickly moving through aisles. &
User’s gaze is shifting frequently; visual queries may be missed. Audio is preferred. &
Offer to recite the user's grocery list &
Binary & Audio \\
\midrule
User is alone in a new restaurant, unfamiliar with the menu. The space is quiet and not crowded. &
User may need help deciding what to order and may benefit from both visual and audio support. &
Provide dish recommendations (\eg, “Top dishes,” “Vegetarian options,” “What I had last time”). &
Multi-choice & Audio + Visual \\
\bottomrule
\end{tabular}
\normalsize
\caption{Representative Few-Shot Examples for Context-Conditioned Query Reasoning}
\vspace{-2em}
\label{tab:few-shot}
\end{table*}
\normalsize

\textbf{Query Format Stability within Task Categories}
We analyzed query type consistency across contextual variants of each activity. A participant’s query type was considered consistent for a task if it appeared in at least 80\% of the contextual variants for that activity. Out of 240 possible activity-participant pairs, 23 met this criterion.

This relatively low consistency rate indicates that users do not adopt a one-size-fits-all querying strategy. Instead, they fluidly adjust their preferences depending on the situation—supporting the notion that context-aware query adaptation is critical for proactive systems.

\textbf{Taxonomy of Desired Proactive Agent Actions}
\label{sec:taxonomy}
To systematically analyze participants' free-text responses, we developed a two-layer taxonomy.

At the first level, we identified core \textbf{action categories}, including \textit{Suggest}, \textit{Remind}, \textit{Guide}, \textit{Summarize}, \textit{Automate}, \textit{Visual Augmentation}, \textit{Information Retrieval}, and \textit{Take App Action}. For instance, a participant in a cooking scenario requested, \textit{“Detect my step and display the next one”}, which was categorized as a \textit{Guide} action.

At the second level, we derived \textbf{contextual categories} that modulate these actions, such as \textit{Familiarity-Based}, \textit{Urgency-Based}, \textit{Social Coordination}, \textit{Cognitive Load}, and \textit{Sensory Disruption}. For example, one participant in a restaurant setting asked the agent to \textit{“brighten the menu and read items aloud,”} which combines \textit{Visual Augmentation} with \textit{Sensory Disruption}.

A list of context categories and action types as well as the distribution among annotated data is shown in Appendix~\ref{app:category_list}.
This analysis provided a structured understanding of user preferences, which in turn guided our manual authoring of the representative few-shot prompts used for in-context learning.

\subsubsection{Context-Conditioned Query Reasoning with In-Context Learning}

To generate context-appropriate proactive agent behavior, we implemented an in-context learning pipeline using LLM. Rather than relying on fixed rules or templates, this module adapts to novel situations by conditioning on structured representations of user context. It outputs a proactive agent action, query type, and presentation modality. The querying strategy draws on our dataset of 937 annotated examples and our taxonomy of user preferences (\S\ref{sec:taxonomy}).

\paragraph{Prompting Strategy.}
Each input to the language model consists of three components: (1) a preamble specifying the agent role and the valid output space; (2) a small set of few-shot examples drawn from our annotated pool; and (3) the user’s current task context, expressed as natural language with key situational factors (activity, location, sensory load, social engagement, familiarity, urgency).
Few-shot examples are composed of three fields:

\begin{itemize}
    \item \textbf{Context:} A natural language description combining relevant situational dimensions such as high-level activity (\eg, cooking, commuting), physical location (\eg, kitchen, museum), environmental factors (\eg, noise, crowd density), user engagement (\eg, hands occupied, visually focused), social context (\eg, alone or with others), and task familiarity or urgency (\eg, rushing, first-time visit).
    
    \item \textbf{Reasoning (Chain-of-Thought):} A brief rationale that connects salient context features to interaction considerations—\eg, decision complexity, input/output availability, or social norms. These reasoning lines were manually authored based on design patterns extracted from our user study, and serve as Chain-of-Thought (CoT) scaffolds for the model.

    \item \textbf{Agent Suggestion:} A structured prediction consisting of (a) a description of the agent's suggested action, (b) a query format (binary, multi-choice, or icon), and (c) a presentation modality (audio, visual, or audio+visual).
\end{itemize}

For instance, a complete triplet might include:
\begin{itemize}
    \item \textit{Context:} “User is in a grocery store, browsing aisles alone, holding a shopping cart, and navigating quickly in a noisy, crowded setting.”
    \item \textit{Reasoning:} “Because the user is rushing and both visually and physically engaged, a binary audio prompt reduces interaction load.”
    \item \textit{Agent Suggestion:} Offer to recite the user’s grocery list | Binary | Audio
\end{itemize}

Table~\ref{tab:few-shot} shows representative exemplars and how context cues map to different query policies.

\paragraph{Few-Shot Selection.}
We provide six exemplars, selected for contextual similarity—matching on high-level activity and at least one additional contextual category (e.g., SIID, social engagement, or familiarity). This balanced diversity with input-length constraints. The prompt then concludes with the user's current context, followed by the final task instruction:


\small
\begin{quote}
\texttt{
Based on the context provided above, generate: \newline
(1) a reasoning for your decision, \newline
(2) the recommended agent action, \newline
(3) a query format (binary/multi-choice/icon), and \newline
(4) a presentation modality (audio/visual/audio+visual). \newline
Structure the output as shown in the examples; if the query format is 'multi-choice', provide three distinct options for the agent action.
}
\end{quote}
\normalsize

\paragraph{Runtime Inference and Output Integration.}
At runtime, a lightweight parser composes a context string from scenario tags or simulated sensor values (e.g., engagement, noise level). The LLM returns an \emph{agent action}, \emph{query type}, and \emph{presentation modality}. We apply simple string parsing to extract these fields and pass them to the interface for immediate rendering (\autoref{fig:system_architecture}). This produces differentiated behavior across context variants while keeping latency within our system constraints (Appendix~\ref{app:latency}).

\subsection{Interaction Modality Implementation}

To support unobtrusive interaction in a variety of contexts, we implemented four user input modalities for responding to the agent: voice, head gestures, hand gestures, and gaze. 
All interaction methods were developed using WebXR\footnote{WebXR APIs: \url{https://immersiveweb.dev}} and three.js\footnote{three.js: \url{https://threejs.org}}, and run in Chrome v137 on an Android XR headset.


\subsubsection{Verbal Interaction}

We implemented verbal input using the Web Speech API, specifically the `SpeechRecognition' interface. The recognition system operates in a limited vocabulary mode, accepting discrete responses such as \texttt{YES}, \texttt{NO}, \texttt{ONE}, \texttt{TWO}, and \texttt{THREE}, corresponding to binary and multi-choice prompts. To improve recognition accuracy and avoid partial matches within longer phrases, we applied regular expression boundaries (\eg, \verb|\bONE\b|). Recognition is filtered by a confidence threshold of 0.7 to reduce false positives.

To support NLCS interaction, we trained a lightweight, real-time classifier using Google’s Teachable Machine~\cite{carney2020teachable}. The classifier distinguishes ambient background noise from affirmative and negative NLCS signals. We trained the model using 120 samples of background noise and 30 samples each of affirmative and negative humming. Internally, Teachable Machine uses a transfer-learned version of Google's YamNet audio classification model, which employs the MobileNet architecture~\cite{howard2017mobilenets} and operates on mel spectrogram representations. The trained model was exported to TensorFlow.js~\cite{smilkov2019tensorflow} and integrated into our WebXR-based framework for real-time NLCS detection.

\subsubsection{Head Gestures}

Head-based interaction leverages the user's rotational head pose, estimated from WebXR's head tracking data. For binary queries, we detect nodding (pitch axis) and shaking (yaw axis) based on oscillatory movement patterns. A gesture is confirmed when 3–4 directional reversals are detected within a fixed temporal window, with a per-frame angular velocity threshold of 0.05 radians to filter out noise and micro-movements.

To enable multi-choice selection, we implemented a tilt-based gesture mapping. This tracks the user's absolute head orientation relative to a neutral reference pose. Tilts beyond 0.3 radians to the left or right are interpreted as selecting the first or second option, respectively. A backward tilt exceeding 0.4 radians selects the third option. This mapping provides a hands-free alternative to UI selection, while maintaining a low motor demand footprint (\textit{\eg}, when user is washing hands).

\subsubsection{Hand Gestures}

Hand input is recognized through WebXR's joint tracking API and analyzed using a geometry-based classifier. Our implementation defines five discrete hand gestures: \texttt{ONE}, \texttt{TWO}, \texttt{THREE}, \texttt{THUMBS\_UP}, and \texttt{THUMBS\_DOWN}. Gestures are recognized based on the extension state of individual fingers, computed via vector alignment and extension ratios derived from the relative positions of joint triplets on each finger.

A finger is marked as extended if the dot product between its segment vectors is near 1 (indicating alignment) and the overall extension exceeds 80\% of its normalized length. For example, a \texttt{TWO} gesture is recognized when the index and middle fingers are both extended, while others remain curled. Thumbs up and down are detected via the dot product between the thumb vector and the global up/down direction, based on the palm normal.

To prevent false positives from brief postural noise, gestures must be held for a minimum duration of 1000ms, sampled at 30ms intervals. Gesture state is tracked over time to ensure stable classification before confirming an input.

\subsubsection{Gaze Interaction}

Due to WebXR limitations, we simulated gaze using head orientation to approximate the user's point of regard. UI elements are treated as collidable objects in 3D space, and selection is inferred based on sustained gaze dwell over a selectable target. A selection is confirmed if the user maintains gaze for at least 3.5 seconds, minimizing accidental activation while preserving low-effort interaction. Future research may leverage native OpenXR or Unity for gaze+pinch interaction~\cite{Pfeuffer2017Gaze}.



\begin{figure*}[h]
  \centering
  \includegraphics[width=\textwidth]{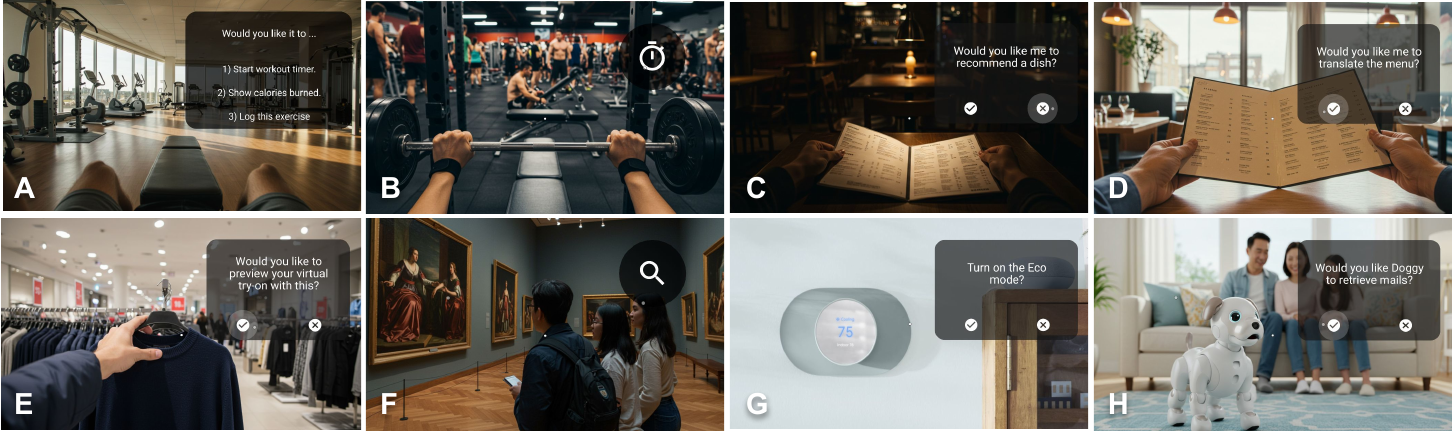}
  \caption{Applications. A-D): \textsc{Sensible Agent}'s initial query (\textit{I}) and repetitive query (\textit{R}), based on the same daily scenarios. A) Gym visit-\textit{I}. B) Gym visit-\textit{R}. C) Restaurant order-\textit{I}. D) Restaurant order-\textit{R}. E) Novel feature suggestion: virtual try-on. F) Subtle cues: information retrieval. G) Effortless smart device control. H) Future application: Human-robot interaction.}
  \label{fig:applications}
\end{figure*}
\section{Applications}

\textsc{Sensible Agent}'s core capability---dynamically adapting \textit{what} proactive assistance to offer and \textit{how} to interact---directly addresses the critical challenge of \textit{interaction friction} that often hinders the practical adoption of proactive AR agents. By intelligently selecting the content and interaction modalities optimized for minimal user efforts and disruption, \textsc{Sensible Agent} enables compelling AR applications that were previously cumbersome or impractical:

\paragraph{\textbf{Context-Adaptive Routine Support}} 
\textsc{Sensible Agent} modifies its behavior based on learned routines, context, and user proficiency.
\begin{itemize}
    \item \textbf{Beginner \textit{vs.} Expert Use}: For a user's very first gym visit (\autoref{fig:applications}A), the system offers multi-choice for assistance, facilitating user exploration. However, for subsequent routine workouts, particularly in a noisy environment where voice input is difficult, it could adapt to allow a simple gaze gesture at an icon to start/stop a timer (\autoref{fig:applications}B), minimizing disruption and leveraging a modality suitable for the context.
    \item \textbf{Learned Preferences}: During a first-time restaurant visit (\autoref{fig:applications}C), the system might proactively offer dish recommendations. If the user dismisses this (\textit{e.g.}, via a head shake) and explicitly requests menu translation, \textsc{Sensible Agent} can infer this preference. On subsequent visits to similar venues (\autoref{fig:applications}D), it can prioritize offering translation assistance proactively, adapting the content of its assistance based on interaction history.
\end{itemize}

\paragraph{\textbf{Opportunistic Suggestion}}

Beyond adapting existing interactions, \textsc{Sensible Agent} can opportunistically introduce users to relevant but unexplored system capabilities. For example, during clothes shopping (\autoref{fig:applications}E), a user might primarily use the AR agent for price checks or list management. \textsc{Sensible Agent} can monitor user activity and, during moments inferred as lower-urgency browse, proactively suggest a related but unused feature, such as virtual try-on. This facilitates feature discovery at moments when the user is likely receptive, without interrupting focused tasks.

\paragraph{\textbf{Minimally Intrusive Augmentation}}

\textsc{Sensible Agent} framework preserves social flow by minimizing interference with the real-world procedures that they augment. In a museum setting (\autoref{fig:applications}F), users might be engaged in conversation with companions. \textsc{Sensible Agent} can provide access to supplementary information (\textit{e.g.}, details about a painting via a subtle search icon) in a manner that requires minimal overt interaction, allowing users to access digital information without significantly disrupting the primary social activity.

\paragraph{\textbf{Potential Extensions: Cross-Device Orchestration}}

As sensing capabilities improve and AR hardware becomes more integrated, we envision \textsc{Sensible Agent} acting as an orchestration engine. Future work could explore how the framework could dynamically select the most appropriate device and modality for a given task – potentially leveraging nearby surfaces as displays, integrating data from smart home sensors (\autoref{fig:applications}G), or coordinating actions with physical robotic agents (\autoref{fig:applications}H) – based on inferred user needs and context, further reducing interaction friction in complex, multi-device environments.

\section{Preliminary Evaluation}
We conducted a within-subjects preliminary evaluation to compare \textsc{Sensible Agent} against a baseline voice-controlled agent, modeled after existing systems like Project Astra~\cite{GoogleDeepMind_ProjectAstra_2024}. This study aimed to surface insights into interaction efficiency, cognitive effort, and user preference across contextually varied scenarios. 

We examined whether \textsc{Sensible Agent} would (1) reduce users’ perceived cognitive load compared to explicit voice-based querying, (2) result in slower overall interaction time due to its two-step confirmation mechanism, and (3) be preferred in repetitive, context-varying situations where users experience situational impairments or fluctuating input/output availability. We also explored whether participants would default to familiar input modalities or adapt their interaction strategies based on context.

\subsection{Participants}
We recruited 10 participants (6 male, 4 female), aged 24--36 ($\mu$ = 30.6), from within our organization via internal mailing lists. All had prior experience using AR headsets ($\mu$ = 3.8) and moderate familiarity with voice-based assistants ($\mu$ = 3.6) on a 5-point Likert scale.

\begin{figure*}[h]
  \centering
  \includegraphics[width=\linewidth]{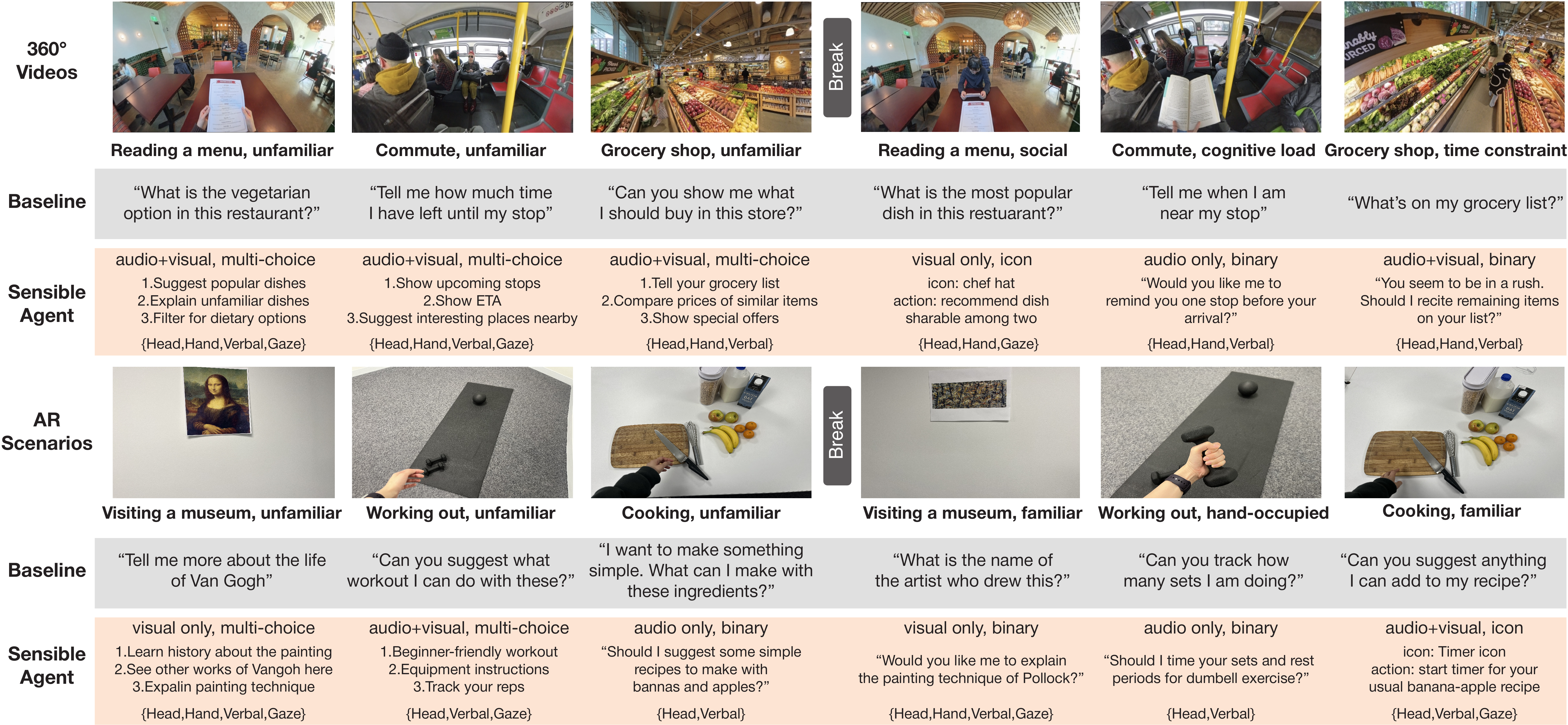}
    \caption{
    Experiment flow across six high-level activities. The top row shows screenshots from the three 360° video scenarios (reading a menu, commute, grocery shopping), and the bottom row depicts physically staged AR scenes (visiting museum, working out, cooking). Each scenario is labeled with its high-level activity and context variant (\eg, \textit{unfamiliar}, \textit{social setting}). Participants experienced all \textit{unfamiliar} scenarios first, followed by their corresponding context variants, avoiding back-to-back repetition of the same activity to simulate naturalistic task switching. For each scenario, we include an example from the baseline condition (user-issued voice query) and a \textsc{Sensible Agent} response, showing the system-selected query type (icon, binary, or multi-choice), presentation modality (audio, visual, or audio+visual), a condensed version of the system-generated prompt, and the available input modalities based on context.
    }

  \label{fig:userstudy-design}
\end{figure*}

\subsection{Apparatus and Experiment Design}
The study was implemented on Project Moohan, an Android XR headset\footnote{Android XR: \url{ https://www.android.com/xr}} using a WebXR-based prototype on Chrome v137. Participants experienced both AR (with video see-through) and VR environments and interacted with AI agent using gaze, voice, hand, and head gestures. A custom WebSocket-based control interface allowed the experimenter to manage the experimental flow remotely, including switching between AR and 360° video environments without requiring headset removal.

The experimental design overview is illustrated in~\autoref{fig:userstudy-design}.
Participants completed twelve scenarios across two system conditions: (1) \textsc{Sensible Agent}, which provided proactive assistance with unobtrusive, multimodal interactions; and (2) a \textit{baseline} system that required users to initiate requests via voice commands, following a conventional assistant model (\eg, \textit{“What should I order?”}, \textit{“Tell me about this exhibit.”}). The system was Wizard-of-Oz controlled;\footnote{The WoZ control was limited to system activation only. All core agent logic, including context sensing and query generation, was fully functional.} participants tapped the headset to signal readiness, after which the experimenter triggered the appropriate system behavior (speech detection for baseline, environment detection for \textsc{Sensible Agent}).

Scenarios were divided into six high-level activities: three delivered as 360° videos (\eg, reading a menu at a restaurant, shopping at a grocery store, commuting in bus), and three physically staged AR scenes (\eg, cooking at a kitchen, working out at a gym, visiting the museum). 
Each activity was experienced in two forms: a \textit{baseline unfamiliar version} and a \textit{context variant} that modulated key context components that were explored in the workshop study and the data collection study, including social engagement, temporal urgency, or sensory constraints. For instance, a ``reading a menu at a restaurant'' scenario may be experienced first alone and then in a social setting. This paired design is to examine how interaction preferences shift when the same high-level activity occurs under altered contextual pressures.

The overall scenario flow was intentionally structured to avoid back-to-back repetitions of the same activity. As shown in ~\autoref{fig:userstudy-design}, participants first experienced three unfamiliar scenarios across different activities. After a short break, they encountered the corresponding three variant scenarios. This design aimed to minimize repetition effects and maintain participant engagement across tasks, while enabling direct within-subject comparison of behavior across unfamiliar and variant contexts.

\begin{figure*}
\scalebox{0.9}{
    \centering
    \begin{subfigure}{0.16\linewidth}
        \centering
        \includegraphics[width=\linewidth]{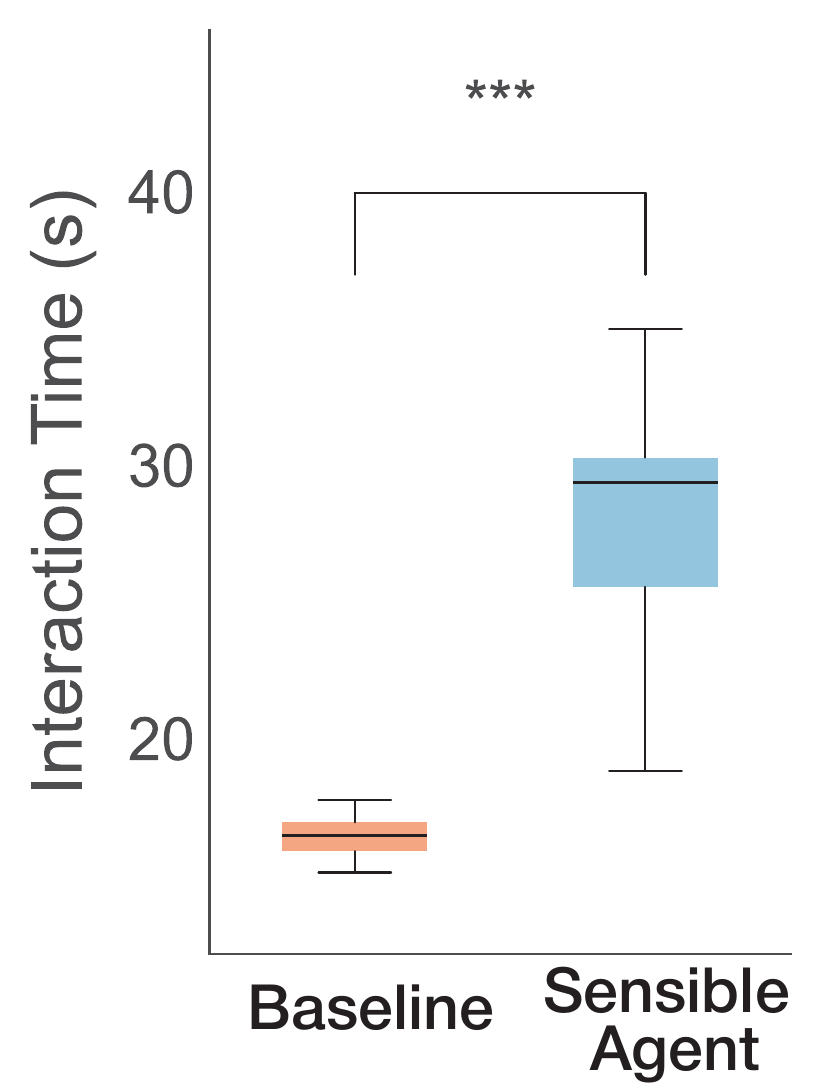}
        \caption{ Interaction Time
        }
        \label{fig:interaction-time}
    \end{subfigure}
    \begin{subfigure}{0.61\linewidth}
        \centering
        \includegraphics[width=\linewidth]{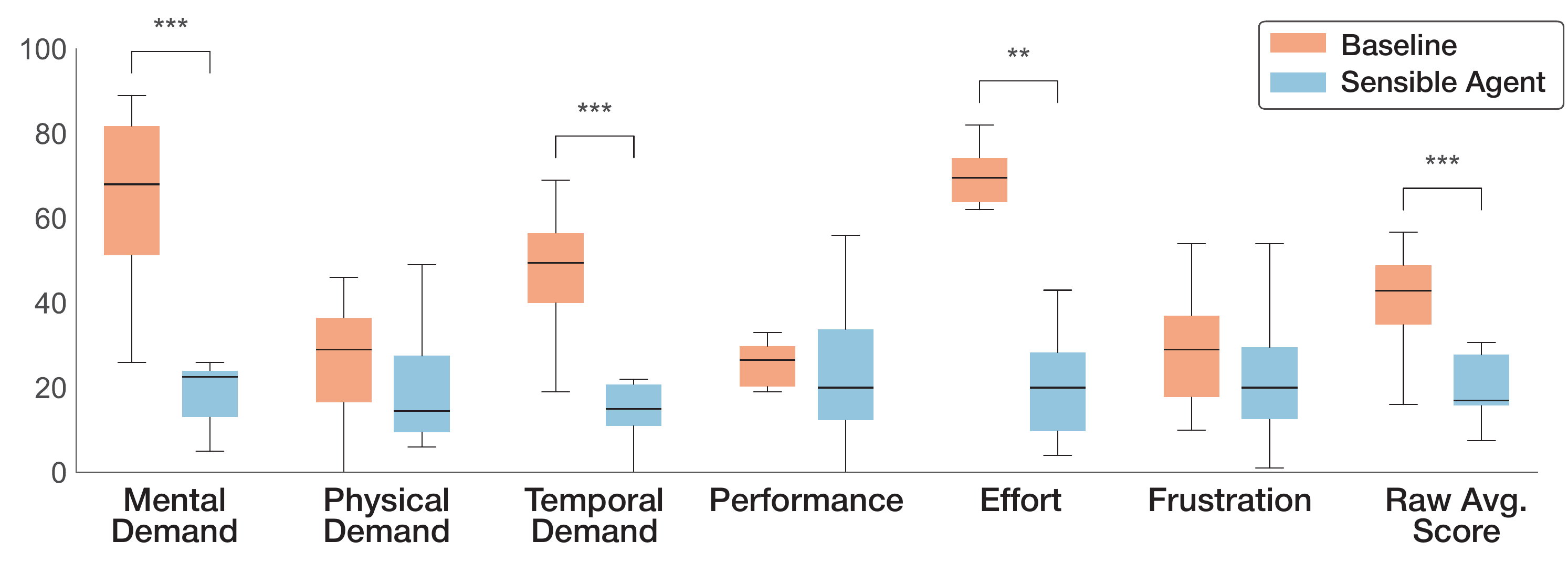}
        \caption{Raw TLX Score}
        \label{fig:nasatlx}
    \end{subfigure}
    \hfill
    \begin{subfigure}{0.15\linewidth}
        \centering
        \includegraphics[width=\linewidth]{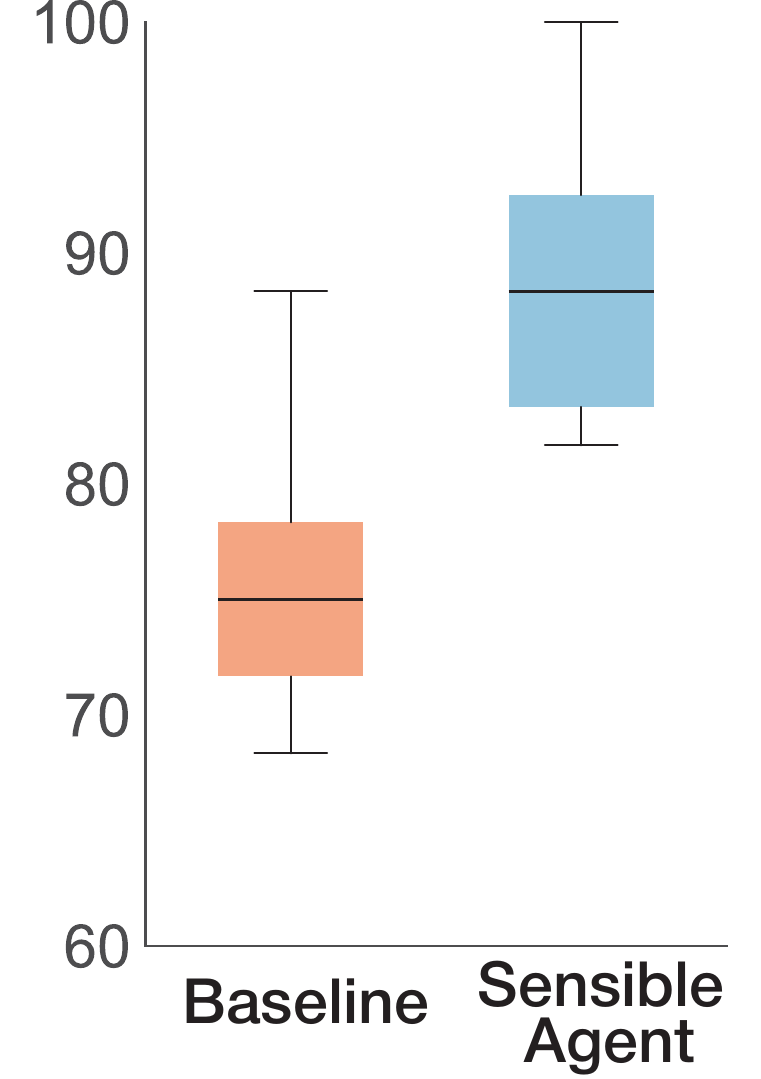}
        \caption{SUS Score}
        \label{fig:sus}
    \end{subfigure}
    \hfill
    \begin{subfigure}{0.14\linewidth}
        \centering
        \includegraphics[width=\linewidth]{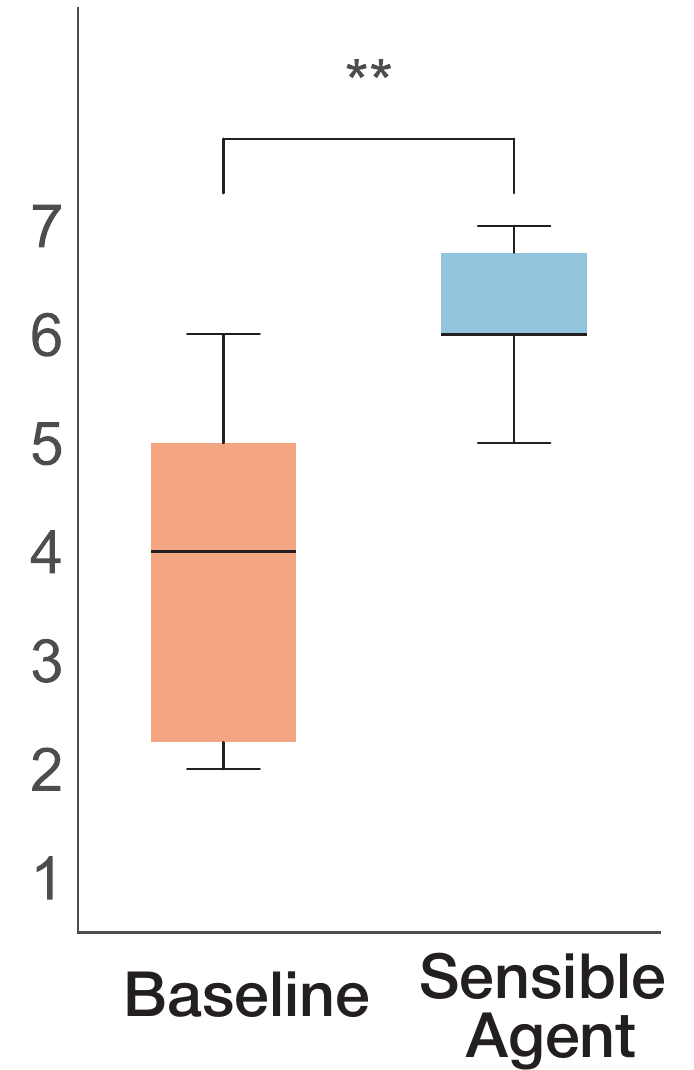}
        \caption{Preference  
        }
        \label{fig:preference}
    \end{subfigure}
    \hfill
    
    }
    \caption{Quantitative analysis of (a) interaction time, (b) Raw TLX scores, (c) SUS scores, and (c) preference measures in our user study. The statistic significance is annotated with $^*$, $^{**}$, or $^{***}$ (representing $p<.05$, $p<.01$, and $p<.001$, respectively).}
    \label{fig:stat-results}
\end{figure*}

\subsection{Procedure}

Each study session lasted approximately 45 minutes. Following a short tutorial on the interaction modalities supported in the system (\eg, head gestures, hand gestures, gaze, or NLCS), participants completed six scenarios in total; three \textit{unfamiliar} baseline versions followed by their paired \textit{context variants}, each drawn from a different high-level activity (\eg, cooking, commuting). To minimize learning or carryover effects, participants never encountered both versions of the same activity back-to-back. The scenario order was consistent across participants and is shown in ~\autoref{fig:userstudy-design}.

At the beginning of each trial, participants were briefly narrated the contextual framing of the scenario, including relevant conditions such as task familiarity, social engagement, or time pressure. During each scenario (lasting ~2–3 minutes), participants responded to agent prompts using the available input modalities. In \textsc{Sensible Agent} conditions, the system adaptively enabled modalities based on context, and participants were free to respond using any that were available. Participants interacted with the agent until a system response was completed.

System conditions were counterbalanced across participants using a Latin square. After each system condition, participants completed the NASA-TLX~\cite{hart1988development} and System Usability Scale (SUS)~\cite{brooke1996sus} questionnaires. We also logged interaction time (from prompt trigger to system response) and recorded the modality used for each response.

At the end of the entire session, participants in a brief semi-structured interview. We asked about overall system preferences, perceived effort, comfort with interaction modalities, and expectations for proactive agent behavior in everyday contexts.




\subsection{Results}

We report both parametric (paired-sample t-test) and non-parametric (Wilcoxon signed-rank test) results, based on Shapiro-Wilk normality checks. All quantitative analyses are exploratory and not corrected for multiple comparisons. Effect sizes (Cohen’s $d$ or rank-biserial $r$) are included for transparency (See \autoref{fig:stat-results}).

\paragraph{\textbf{Interaction Time}}
Interaction was faster in the baseline voice-query condition ($\mu$ = 16.43s, $\sigma$ = 0.84) than in the \textsc{Sensible Agent} condition ($\mu$ = 28.54s, $\sigma$ = 4.85), $t(9) = -7.54$, $p < .001$, $d = -2.51$. This trend is expected due to \textsc{Sensible Agent}’s two-step interaction flow, where the system first presents a suggested query based on context and the user then confirms or modifies it, in contrast to the baseline system where users immediately issue a voice command.

\paragraph{\textbf{Cognitive Load (NASA-TLX)}}
\textsc{Sensible Agent} showed lower mental demand ($\mu$ = 21.10, $\sigma$ = 11.57) than the baseline ($\mu$ = 65.00, $\sigma$ = 20.19), $t(9) = 6.40$, $p < .001$, $d = 2.03$. Participants also reported lower temporal demand with \textsc{Sensible Agent} ($\mu$ = 16.00, $\sigma$ = 10.12) versus baseline ($\mu$ = 46.20, $\sigma$ = 16.61), $t(9) = 5.43$, $p < .001$, $d = 1.72$. 

Effort scores showed a similar trend: $W = 1.00$, $p = .0039$ (Wilcoxon), with \textsc{Sensible Agent} rated lower ($\mu$ = 20.30, $\sigma$ = 12.79) than baseline ($\mu$ = 67.20, $\sigma$ = 14.70). Participants frequently mentioned the ease of interaction as a key factor. Consistent differences were observed for physical demand ($p = .18$), performance satisfaction ($p = .65$), or frustration ($p = .23$).

Total Raw-TLX scores were lower for \textsc{Sensible Agent} ($\mu$ = 20.55, $\sigma$ = 8.42) than baseline ($\mu$ = 43.27, $\sigma$ = 9.73), $t(9) = 6.76$, $p < .001$, $d = 2.14$, reflecting a consistent pattern of reduced cognitive burden across conditions.

\paragraph{\textbf{Usability (SUS)}}
SUS scores were calculated using the standard procedure ($\sum \text{item scores} \times 2.5$, yielding a range of $[0, 100]$). There was no observed difference in SUS scores between the baseline ($\mu$ = 76.67, $\sigma$ = 5.93) and \textsc{Sensible Agent} ($\mu$ = 81.33, $\sigma$ = 6.58), $W = 11.00$, $p = .11$. Both conditions averaged above 71.4, which maps to a ``Good” usability rating~\cite{bangor2009determining}. Detailed subscale scores are provided in Appendix~\ref{app:sus_sub}.

\paragraph{\textbf{User Preferences}}
In a 7-point Likert scale, participants expressed a preference for \textsc{Sensible Agent} ($\mu$ = 6.00, $\sigma$ = 0.94) over the baseline ($\mu$ = 3.80, $\sigma$ = 1.48), $W = 0.00$, $p = .0074$. Seven of ten participants expressed that the proactive and unobtrusive nature of \textsc{Sensible Agent} made interaction more engaging. 

\paragraph{\textbf{Interaction Patterns.}}
We observed notable patterns in how participants interacted with the multi-choice panel across different context scenarios. In the initial round of unfamiliar scenarios, those presented first in both the 360° and AR settings, participants selected different input modalities when prompted to confirm agent suggestions: 3 participants used voice, 2 used head gestures, and 4 used hand input. Six out of ten participants consistently used the same modality throughout this round, suggesting an early anchoring effect in input behavior. The remaining four participants changed modalities during the round, often out of exploratory intent. 

Situational factors influenced modality choice. In scenarios where participants were physically holding objects—such as the gym (holding a dumbbell) or cooking (manipulating ingredients)—six participants switched to hands-free modalities like head or voice input, indicating sensitivity to situational input constraints (SSID). Two participants preferred head gestures, citing familiarity with similar gestures on consumer devices like Apple's AirPods. However, they noted that the tilting gesture required some adaptation. Gaze-based input was met with mixed responses; five participants reported that gaze sometimes conflicted with reading the panel content, leading them to look away intentionally to avoid accidental selections.

\subsection{Qualitative Feedback}
We summarize key takeaways from post-trial interviews, focusing on user perceptions of input, prompt format, and interactional fluency.

\textbf{Preference for unobtrusive input.} Five participants (P1, P2, P5, P8, P9) appreciated being able to respond using minimal-effort inputs such as head gestures or short verbal confirmations. P2 noted, \textit{“I liked how I could reply in almost any way I wanted.”} P9, \textit{“It works well because the agent asked me something ... I didn’t even have to think to ask. It was natural to just nod and keep going.”} P1 highlighted the benefit of quick interactions: \textit{“I could answer fast and move on. That’s what makes it feel helpful instead of annoying.”} P8 added, \textit{“I loved how little effort I had to give to respond, which makes even more sense in situations where I have a friend who wants me to focus on them during a conversation.”} Three participants (P3, P6, P10) expressed a desire for lightweight confirmation that their input had been successfully recognized.

\textbf{Prompt format preferences.} All ten participants found multi-choice prompts especially helpful in unfamiliar situations. P4 stated, \textit{“The choices were helpful when I didn’t know what I want. I could just pick.”} Opinions on icon-only formats were mixed; while some appreciated their brevity, P7 mentioned interpretability challenges: \textit{“It took me three seconds to figure out what the icon meant.”}

\textbf{Alignment with social interaction patterns.} Several participants (P3, P8, P10) described the interaction style as resembling a casual conversation, where suggestions were context-aware and effort-free. As P10 put it, \textit{“It felt like just talking to someone who already knows what I might want.”} 

\section{Discussion}

Our study reveals that proactive agents, when equipped with unobtrusive multimodal interfaces, not only reduce user effort but also reshape how users perceive and engage with digital assistants. We reflect on these broader implications and discuss how our framework can evolve based on observed behaviors from both studies.

\subsection{Proactivity as a Social Cue}
While prior work on proactive agents has focused on reducing friction or predicting user needs, our findings suggest that proactivity may also shape how users perceive the agent as a social presence. Seven out of ten participants reported that they found interactions with \textsc{Sensible Agent} to be more engaging or even enjoyable beyond simply requiring less effort. 

The use of subtle, non-verbal input methods, such as nodding or tilting the head, further contributed to this perceived naturalness. Participants likened these gestures to the kinds of acknowledgments they use in everyday social interactions. 
These results point to a broader design opportunity: proactively adaptive systems may benefit from aligning more closely with human social cues, not only to reduce effort, but to foster rapport and interactional fluency.

\subsection{Modality Fusion Based on Situational Weighting}

While our current framework allows users to respond using multiple modalities, input signals are handled independently and without coordination. During our workshop study, two expert participants proposed that the system could go further by integrating multiple modalities using situational weighting. For instance, in real-world environments, users may emit overlapping or even conflicting cues—such as unintentionally nodding while verbally responding ``\textit{no}'', or glancing away while affirming an option aloud. Such ambiguities are difficult to resolve without considering environmental context, task state, and historical user behavior.

This insight suggests an extension to our framework where each input modality is evaluated not only by its raw signal, but by its reliability under the current situational context. For example, voice input may be down-weighted in noisy environments, while hand gestures may be deprioritized when the user is physically constrained. This would allow the agent to compute a weighted confidence score across inputs, resulting in more robust intent inference. Building in such a multimodal arbitration layer—responsive to situational impairments and social context—would position proactive agents to operate more effectively in complex, dynamic environments.

In safety-critical or task-oriented contexts such as equipment repair or medical triage, such multimodal arbitration becomes essential—for both robustness and to modulate intrusiveness and timing. Our modular architecture could incorporate a task monitor that tracks procedural stages and interruptibility signals (\eg, idle hands, pause in activity), enabling the agent to shift from ambient assistance to structured, stepwise support. This would allow proactive prompting to align not just with user availability, but with task flow.

\subsection{Modeling Modality Preferences}

While \textsc{Sensible Agent} currently treats each proactive interaction as a discrete event, several participants attempted to engage in follow-up utterances or gestures, suggesting a desire for extended, multi-turn exchanges. This highlights an opportunity to extend our framework by incorporating a temporal layer that tracks dialogue state and user responsiveness across turns.

One open design question is how agents should adapt presentation strategies during sustained interactions. Should the same modality persist across turns to support continuity and reduce surprise, or should the system adjust dynamically to maintain engagement or match user behavior? Extending our framework to include a temporal rhythm model could enable systems to better scaffold ongoing interactions—particularly in context-rich environments where attention and modality availability fluctuate. 

In addition to managing temporal rhythm, agents must also account for longer-term user preferences that may not align with situationally optimal choices. For instance, a user might consistently prefer hand gestures over voice even in quiet environments. Our framework's notion of a \textit{user action prior} (\autoref{fig:framework} already supports situational preference modeling; future extensions could incorporate longer-term adaptation through techniques like reinforcement learning or implicit feedback tracking. This would enable the system to personalize modality strategies based not just on immediate context, but on evolving user tendencies and habits.

\section{Limitations}

Our current prototype and study were designed to validate core interaction principles; however, several areas remain open for future extension and evaluation. First, while our design supports personalization and contextual adaptation, our current prototype does not model user history or longitudinal preferences. Incorporating historical interaction patterns—such as preferred modalities, timing preferences, or context-specific behaviors—could enable more personalized and anticipatory agent behavior over time.

Second, we did not model the precise timing of proactive prompts within a given situation. Prior work in procedural task guidance has explored step-aware interventions where action boundaries are clearly defined~\cite{li2025satori, bandyopadhyay2025yeti}. However, our target contexts involved open-ended activities (\eg, browsing a museum, ordering food) where task state is fluid and interruption thresholds are less well-defined. Determining the optimal timing for intervention in such scenarios remains an open challenge and a promising extension to our framework.

Third, while our study compared  to a conventional voice-query baseline, we did not include a condition where assistance was delivered through always-visible multi-choice interfaces using conventional XR interactions (\eg, point + pinch) or microgestures (\eg, gaze + thumb-to-finger swipe~\cite{Pei2024UI}). Comparing against such persistent UI paradigms could help isolate the specific contribution of unobtrusive, gaze- or head-based modalities to user experience and perceptions of agent presence. Additionally, our sample size was limited (n=10), and the findings should be interpreted as preliminary and exploratory. While we observed consistent trends, future work should expand to assess generalizability and long-term effectiveness.

Finally, the scenarios used in our study focused on everyday, repeated contexts—such as grocery shopping or exercising—but did not include task-oriented settings with subsequent steps to take or high-stakes outcomes. Future work could explore how proactive, multimodal agents function in domains such as collaborative work, task guidance, or healthcare, where expectations and risks differ.

\section{Conclusion and Future Work}
In this paper, we address the critical challenge of interaction friction hindering the adoption of proactive AR agents. Existing approaches often rely on explicit, burdensome interactions unsuitable for many real-world contexts. Our work introduced \textsc{Sensible Agent}, a framework demonstrating the feasibility and benefit of dynamically adapting both the content (“what”) and modality (“how”) of proactive assistance to achieve unobtrusive interaction. By leveraging multimodal sensing and LMM reasoning, \textsc{Sensible Agent} selects contextually appropriate actions and interaction methods designed to minimize user effort. Our evaluation confirmed that this dynamic adaptation significantly reduces perceived intrusiveness and interaction burden compared to conventional methods, paving the way for more seamless human-AI collaboration in AR.

While \textsc{Sensible Agent} represents a significant step towards effortless proactive AR, several avenues warrant further exploration. Future work should focus on expanding the repertoire of unobtrusive interaction modalities beyond the current set (\textit{\eg}, subtle haptics, ambient visualizations) and exploring adaptation within modalities (\textit{\eg}, varying the level of detail). Developing rigorous benchmarks and standardized metrics to evaluate the effectiveness and unobtrusiveness of proactive AR agents remains a key challenge for the community. Furthermore, extending the framework to provide more personalized, long-term recommendations based on inferred user goals and preferences presents an exciting direction. Finally, integrating \textsc{Sensible Agent}'s principles into broader ambient computing~\cite{olwal2022hidden} and mirrored world~\cite{Du2019Geollery} environments could unlock truly pervasive, yet respectful, proactive assistance.

\begin{acks}
We would like to thank Zhongyi Zhou, Vikas Bahirwani, Jessica Bo, Zheng Xu, Renhao Liu for their feedback and discussion on our early-stage proposal. We thank Alex Olwal for founding and directing the Interaction Lab at Google Research, directing the Augmented Language workstream, and pioneering research in \cite{olwal2009unobtrusive} that inspired this work.
\end{acks}

\bibliographystyle{ACM-Reference-Format}
\bibliography{_reference}
\newpage
\appendix
\section{Action and Context Categories}
\label{app:category_list}
This section details on the analysis of the action and context categories derived from our formative data collection study (see main paper, Section~\ref{sec:dcs}). The context variants were pre-defined by the scenarios presented to participants, while the action categories emerged from a thematic analysis of the collected user responses.

\subsection{Context Categories}

The context categories were systematically designed into our study scenarios to elicit proactive AI behaviors under various situational constraints. Each scenario was grounded in a high-level activity: commuting on a bus, cooking in a kitchen, working out at a gym, reading a menu at a restaurant, browsing a museum, or grocery shopping. The specific contextual variants for these activities were:
\textbf{Default}: The baseline version of the high-level activity with no specific contextual impairment or modulation. For example, in the `restaurant' scenario, the user is simply reading a menu.

\textbf{Temporal Urgency}: The user faces a time constraint or pressure. For example, the user has to order quickly at the restaurant before a movie starts.

\textbf{Familiarity-Based}: The user is familiar with the environment or task. For example, the user is reading a menu at a restaurant they visit often.

\textbf{Unfamiliarity-Based}: The user is new to the environment or task. For example, the user is at a new restaurant in town for the first time.

\textbf{Cognitive Load}: The user is mentally or physically occupied with a secondary task. For example, reading a book while commuting on the bus or holding dumbbells while working out (hands-occupied).

\textbf{Crowded}: The surrounding environment is noisy and populated with other people. For example, the restaurant is busy and filled with patrons.

\textbf{Socially-Engaged}: The user is directly interacting with another person during the primary activity. For example, talking to a friend while deciding what to order at the restaurant.

\textbf{Divergent Setting (diff)}: The user performs the same high-level activity but in a different type of venue, which may alter their needs or interaction patterns. For example, ordering from a casual café instead of a formal restaurant.

\textbf{Environmental Changes (env)}: The immediate physical environment is altered in a way that creates a sensory challenge. For example, the user is ordering from a restaurant with dim lighting or trying to read an outdoor menu while it is raining.

The Crowded and Socially-Engaged categories introduce social constraints that an agent must navigate. Also, certain scenarios within Cognitive Load (e.g., hands-occupied) and Environmental Changes (e.g., dim lighting) represent forms of Sensory and Situational Impairments/Disabilities (SSIDs), where the user’s ability to interact with standard interfaces is temporarily limited.

\subsection{Action Categories}

From the 937 user-generated responses, we identified eight distinct categories of proactive actions that users desired from the agent. These categories are defined as follows:

\textbf{Suggest}: Proposes a set of options or a single recommendation to aid the user's decision-making process (e.g., ``Recommend a popular dish").

\textbf{Remind}: Surfaces timely, contextually relevant information that the user may have forgotten or needs to be aware of (e.g., ``Remind me to get off at my stop two stops beforehand").

\textbf{Guide}: Provides turn-by-turn or step-by-step instructions to help the user complete a process (e.g., ``Show me the step-by-step instructions on how to make this recipe").

\textbf{Summarize}: Processes a larger body of information to generate a new, condensed version. (e.g., "Summarize the description of this painting for me.").

\textbf{Automate}: Executes a pre-defined, multi-step workflow, often chaining together multiple actions that would otherwise need to be done manually. (e.g., ``Log today's workout.").

\textbf{Visual Augmentation}: Augmenting mixed-reality overlays directly onto the physical world to enhance the user's perception (e.g., ``Highlight the gluten-free items on the menu").

\textbf{Information Retrieval}: Fetches and presents a single, discrete piece of existing data without altering it. (e.g., in a museum, ``Tell me who the artist is for this art piece").

\textbf{Take App Action}: Executes a single, explicit command within a software application. It is a direct trigger for one function. (e.g., ``Play my workout playlist on Spotify").

\begin{table}
\centering
\begin{tabular}{llr}
\toprule
\textbf{Action Category} & \textbf{Context Category} & \textbf{Count} \\
\midrule
Automate & Default & 6 \\
Automate & Cognitive Load & 13 \\
Automate & Familiarity-Based & 46 \\
Automate & Unfamiliarity-Based & 10 \\
Automate & Divergent Setting & 11 \\
Automate & Temporal Urgency & 13 \\
\midrule
Guide & Default & 11 \\
Guide & Unfamiliarity-Based & 34 \\
Guide & Divergent Setting & 9 \\
Guide & Environmental Changes & 17 \\
\midrule
Information Retrieval & Default & 7 \\
Information Retrieval & Unfamiliarity-Based & 39 \\
Information Retrieval & Divergent Setting & 22 \\
Information Retrieval & Environmental Changes & 6 \\
Information Retrieval & Temporal Urgency & 18 \\
\midrule
Remind & Default & 26 \\
Remind & Cognitive Load & 18 \\
Remind & Familiarity-Based & 21 \\
Remind & Divergent Setting & 21 \\
Remind & Temporal Urgency & 33 \\
\midrule
Suggest & Default & 43 \\
Suggest & Cognitive Load & 31 \\
Suggest & Familiarity-Based & 41 \\
Suggest & Unfamiliarity-Based & 14 \\
Suggest & Socially-Engaged & 134 \\
Suggest & Crowded & 52 \\
Suggest & Divergent Setting & 23 \\
Suggest & Temporal Urgency & 30 \\
\midrule
Summarize & Default & 4 \\
Summarize & Cognitive Load & 6 \\
Summarize & Temporal Urgency & 4 \\
\midrule
Take App Action & Default & 24 \\
Take App Action & Cognitive Load & 12 \\
Take App Action & Socially-Engaged & 42 \\
Take App Action & Divergent Setting & 24 \\
Take App Action & Temporal Urgency & 13 \\
\midrule
Visual Augmentation & Default & 3 \\
Visual Augmentation & Crowded & 47 \\
Visual Augmentation & Environmental Changes & 16 \\
Visual Augmentation & Divergent Setting & 5 \\
\bottomrule
\end{tabular}
\caption{Distribution of desired proactive actions (N=937) across eight context categories and eight action categories}
\label{tab:action_context_counts}
\end{table}

\subsection{Context Variant, Query Type, Activity Distribution}
To provide a comprehensive overview of our dataset (N=937), we analyzed the distribution of user responses from two primary perspectives.

First, Table~\ref{tab:action_context_counts_final_v2} details the complete breakdown of desired proactive behaviors. It cross-references the eight action categories (e.g., Suggest, Guide) with the ten contextual variants (e.g., Temporal Urgency, Cognitive Load), illustrating which specific actions were requested most frequently in each situation.

\begin{figure}[h]
  \centering
  \includegraphics[width=\linewidth]{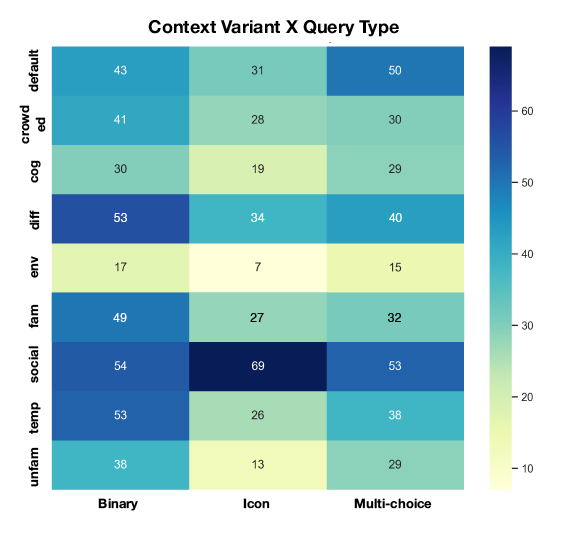}
  \caption{Distribution of response counts across different contextual variants (rows) and query types (columns).}
  \label{fig:dist-count}
\end{figure}

Second, Figure~\ref{fig:dist-count} visualizes the distribution of response counts across the different contextual variants (rows) and query types (columns). The data shows that preferences for query formats shift based on the user's situation. For example, higher counts are observed for binary-choice and multiple-choice formats in variants such as Divergent Setting, Socially-Engaged, and Temporal Urgency. A notable peak is observed for the combination of Socially-Engaged contexts and icon-based queries, which accounted for 69 entries.

\subsection{Usefulness Rating}
\begin{figure}[h]
  \centering
  \includegraphics[width=\linewidth]{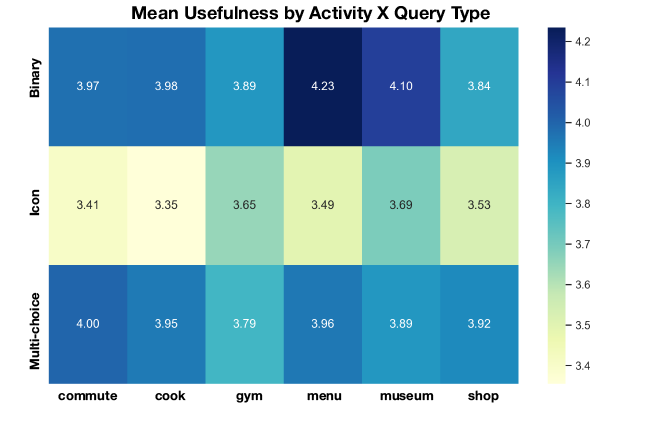}
  \caption{Heatmap showing the mean usefulness ratings of query responses across combinations of activity type (horizontal axis) and query type (vertical axis). Darker shades represent higher perceived usefulness, with binary queries during menu and museum activities receiving the highest ratings.}
  \label{fig:dist-use}
\end{figure}

As shown in Figure~\ref{fig:dist-use}, the perceived usefulness of query responses varies by both activity context and query format.

\section{System Latency and Prompting Efficiency}
\label{app:latency}
To assess the responsiveness of our system, we measured the agent-side latency of the Sensible Agent pipeline—from the moment a new context is constructed to the moment a complete proactive prompt (including reasoning, action type, query format, and presentation modality) is returned from the language model. This evaluation focuses specifically on generation time and does not include user response latency or downstream input handling.

We conducted 20 trials using diverse test contexts spanning multiple high-level activities (e.g., grocery shopping, museum visit, working out) and context variants (e.g., time pressure, social setting, sensory impairment). Each trial used a prompt containing 6 few-shot exemplars selected based on contextual similarity heuristics, such as shared engagement type or environmental condition. The prompts were structured in a triplet format (context, reasoning, agent suggestion), and processed through GPT-4o via API.

Average generation latency was \textbf{6.2 seconds} ($\sigma = 0.8$) on a MacBook Pro (M1 Pro, 16GB RAM) with a stable internet connection. We observed minor variation based on the number and length of examples, but no critical delays for short-to-medium interactions (2–4 minutes) as used in our scenarios.

Compared to prior systems that employed structured LLM-based reasoning, our design reflects a trade-off between expressivity and responsiveness. \textit{Human I/O}~\cite{Liu2024Human} reported an average latency of 19.95 seconds using GPT-4 and 7.33 seconds with GPT-3.5, using full Chain-of-Thought prompts focused on SIID detection. \textit{OmniActions}~\cite{li2024omniactions} did not report exact latency, but leveraged fixed-size prompts for classification over a closed action label set. In contrast, our goal was to support open-context prompting with lightweight, interpretable few-shot examples while remaining within an acceptable delay for proactive interaction in everyday mobile settings.

Although our prompt set covers only a subset of the context-action space observed during data collection, we found that GPT-4o generalized well when exemplars shared key behavioral constraints. Future work may explore integrating retrieval-augmented generation or compact local models to further scale coverage while maintaining or improving latency.

\section{SUS Subscale Scores}
\label{app:sus_sub}
We further examined each subscale component of the SUS questionnaire to investigate whether participants perceived differences between the two systems on specific usability aspects. While no subscale reached statistical significance after correction, we report descriptive comparisons to contextualize user preferences. For example, for the item \textit{“I think that I would like to use this system frequently”}, ratings were slightly higher for the Sensible Agent condition ($\mu=4.3$, $\sigma=0.67$) compared to Baseline ($\mu=2.1$, $\sigma=1.2$), $W=10.0$, $p=.09$. Similarly, for the item \textit{“I found the system very cumbersome to use”}, Sensible Agent received higher ratings ($\mu=3.3$, $\sigma=0.48$) than Baseline ($\mu=2.2$, $\sigma=1.0$), $W=12.0$, $p=.15$, suggesting a possible reduction in perceived complexity (Note that negatively worded items were reverse-scored prior to analysis.). No other items showed notable differences (all $p > .2$), including \textit{“I felt very confident using the system”} and \textit{“I would imagine that most people would learn to use this system very quickly”}, which were rated similarly across conditions.

\end{document}